\documentclass[journal]{IEEEtran}

	\usepackage{graphicx}
	
	\graphicspath{{tikz/}{figures/}}
    \usepackage[
    colorlinks=true,
    linkcolor=blue,
    urlcolor=black,
    citecolor=blue,
    ]{hyperref}
    \usepackage{orcidlink}
    \usepackage{adjustbox}  
    \usepackage[font=footnotesize, labelfont=sf, textfont=sf]{subcaption}
    \usepackage{tabularray}

    \usepackage{amssymb}
    \usepackage{pifont}  
    \newcommand{\cmark}{\ding{51}}%
    \newcommand{\xmark}{\ding{55}}%
    \usepackage{siunitx}    
    
	\usepackage{empheq}     
	\usepackage{stfloats}  
	\usepackage{array} 

    \usepackage{multirow}   
    \usepackage{arydshln}   
    \usepackage{threeparttable}
    \usepackage{fancyhdr}

    \usepackage[ruled,vlined]{algorithm2e}

    \usepackage{xifthen}    
    \usepackage{url}

	\usepackage{cite}
    \usepackage[nopostdot,symbols,nogroupskip,nonumberlist]{glossaries-extra} 
    \glsdisablehyper

	\usepackage{color}

    \newcommand{\tens}[1]{\boldsymbol{\mathcal{#1}}} 

    \newcommand{\prox}{\mathop{\mathrm{prox}}}

    \newcommand{\lexi}{\mathop{\mathrm{matr}}}

    \newcommand{\lino}[1]{\mathop{\mathbb{#1}}}   

    \newcommand{\hadam}{\odot}       
    \newcommand{\cnvo}{\;*\;}        
    \newcommand{\fenc}{^\star}       
    \newcommand{\masked}{^{\boxdot}} 

    \newcommand{\ratio}{\rho}       
    \newcommand{\range}[3]
    {%
         \ifthenelse{\equal{#1}{}}{\left[#2,...\,,#3\right]}{#1\in\left[#2,...\,,#3\right]}%
    }

    \definecolor{added}{RGB}{0,0,0}

\usepackage{booktabs}

\glssetcategoryattribute{acronym}{indexonlyfirst}{true}
\newignoredglossary{ignored}

\newabbreviation{hri}{HRI}{high resolution image}
\newabbreviation{lri}{LRI}{low resolution image}
\newabbreviation{gt}{GT}{ground truth}
\newabbreviation{pan}{PAN}{panchromatic}
\newabbreviation{ms}{MS}{multispectral}
\newabbreviation{hs}{HS}{hyperspectral}
\newabbreviation{hsi}{HSI}{hyperspectral imaging}
\newabbreviation{rgb}{RGB}{red-green-blue}
\newabbreviation{rgbw}{RGBW}{red-green-blue-white}
\newabbreviation{cym}{CYM}{cyan yellow magenta}
\newabbreviation{swir}{SWIR}{short wave infrared}
\newabbreviation{nir}{NIR}{near infrared response}
\newabbreviation{vis}{VIS}{visible}
\newabbreviation{uv}{UV}{ultraviolet}
\newabbreviation{ir}{IR}{infrared}
\newabbreviation{gsd}{GSD}{ground sample distance}

\newabbreviation{exp}{EXP}{interpolated image}
\newabbreviation{tps}{TPS}{thin plate spline}
\newabbreviation{idw}{IDW}{inverse distance weighting}
\newabbreviation{rbf}{RBF}{radial basis function}
\newabbreviation{tps-rbf}{TPS-RBF}{\glsentrylong{tps} \glsentrylong{rbf}}
\newabbreviation{gcv}{GCV}{generalized cross validation}
\newabbreviation{sure}{SURE}{Stein's unbiased risk estimate}
\newabbreviation{pchip}{PCHIP}{piecewise cubic Hermite interpolating polynomial}
\newabbreviation{tin}{TIN}{triangulated irregular network}

\newabbreviation{snr}{SNR}{signal to noise ratio}
\newabbreviation{mtf}{MTF}{modulation transfer function}

\newabbreviation{nn}{NN}{nearest neighbour}
\newabbreviation{cs}{CS}{component substitution}
\newabbreviation{mra}{MRA}{multiresolution analysis}
\newabbreviation{glp}{GLP}{generalized Laplacian pyramid}
\newabbreviation{wb}{WB}{weighted bilinear}
\newabbreviation{id}{ID}{intensity difference}
\newabbreviation{sd}{SD}{spectral difference}
\newabbreviation{iid2}{ItID}{iterative intensity difference}
\newabbreviation{isd}{ItSD}{iterative spectral difference}
\newabbreviation{ap}{AP}{alternating projections}
\newabbreviation{msg}{MSG}{multiscale gradients}
\newabbreviation{ri}{RI}{residual interpolation}
\newabbreviation{ari}{ARI}{adaptative residual interpolation}
\newabbreviation{mlri}{MLRI}{minimized-Laplacian residual interpolation}
\newabbreviation{btes}{BTES}{binary tree-based edge-Sensing}

\newabbreviation{bt2}{BrT}{Brovey transform}
\newabbreviation{gihs}{GIHS}{generalized intensity hue saturation}
\newabbreviation{gs}{GSA}{Gram-Schmidt}
\newabbreviation{gsa}{GSA}{Gram-Schmidt adaptive}
\newabbreviation{atwt}{ATWT}{à trous wavelet transform}
\newabbreviation{hm}{HM}{histogram matching}
\newabbreviation{cbd}{CBD}{context based decision}
\newabbreviation{hpm}{HPM}{high pass modulation}
\newabbreviation{mtf-glp}{MTF-GLP}{\glsentryshort{mtf}-matched \glsentrylong{glp}}
\newabbreviation{mtf-glp-cbd}{MTF-GLP-CBD}{\glsentryshort{mtf}-matched \glsentrylong{glp} with context based decision injection}
\newabbreviation{mtf-glp-hpm}{MTF-GLP-HPM}{\glsentryshort{mtf}-matched \glsentrylong{glp} with high pass modulation injection}
\newabbreviation{bdsd}{BDSD}{band-dependent spatial detail}
\newabbreviation{bayn}{BayesNaive}{Bayesian with naive regularization}
\newabbreviation{hysure}{HySure}{hyperspectral superresolution}
\newabbreviation{cnmf}{CNMF}{coupled nonnegative matrix factorization}
\newabbreviation{pca}{PCA}{principal component analysis}

\newabbreviation{psnr}{PSNR}{peak \glsentrylong{snr}}
\newabbreviation{scc}{sCC}{spatial cross-covariance coefficient}
\newabbreviation{ergas}{ERGAS}{relative dimensionless global error in synthesis}
\newabbreviation{sam}{SAM}{spectral angle mapper}
\newabbreviation{q2n}{\ensuremath{Q^2n}}{\ensuremath{Q2^n} index}
\newabbreviation{uiqi}{UIQI}{universal image quality index}
\newabbreviation{ssim}{SSIM}{structural similarity}
\newabbreviation{med}{MED}{mean Euclidean distance}

\newabbreviation{bin}{BIN}{radiometric binning}
\newabbreviation{cassi}{CASSI}{compressive coded aperture spectral imaging}
\newabbreviation{mrca}{MRCA}{multiresolution coded acquisition}
\newabbreviation{jodefu}{JoDeFu}{joint demosaicing and fusion}
\newabbreviation{dmd}{DMD}{digital micromirror device}
\newabbreviation{doas}{DOAS}{differential optical absorption spectroscopy}

\newabbreviation{bt}{BT}{binary tree}
\newabbreviation{ubt}{UBT}{uniform \glsentrylong{bt}}
\newabbreviation{dbt}{DBT}{dominant \glsentrylong{bt}}
\newabbreviation{cove}{COVE}{\glsentrylong{pan} coverage}
\newabbreviation{peri}{PERI}{periodic}
\newabbreviation{diag}{DIAG}{diagonal}
\newabbreviation{vert}{VERT}{vertical}
\newabbreviation{btree}{BT}{binary tree}
\newabbreviation{rand}{RAND}{random pick}
\newabbreviation{mxdis}{MAXDIS}{maximum distance}
\newabbreviation{diri}{DIRI}{Dirichlet distribution based}
\newabbreviation[shortplural={s.v.}]{sv}{s.v.}{singular value}
\newabbreviation{svd}{SVD}{singular value decomposition}
\newabbreviation{pmd}{PMD}{penalized matrix decomposition}
\newabbreviation{l0}{L0}{level 0}
\newabbreviation{l1}{L1}{level 1}
\newabbreviation{l2}{L2}{level 2}
\newabbreviation{l3}{L3}{level 3}

\newabbreviation{pmf}{pmf}{probability mass function}
\newabbreviation{pdf}{pdf}{probability density function}
\newabbreviation{iid}{i.i.d.}{independent and identically distributed}
\newabbreviation{psd}{PSD}{power spectral density}
\newabbreviation{mse}{MSE}{mean square error}
\newabbreviation{rmse}{RMSE}{root \glsentrylong{mse}}
\newabbreviation{afn}{AFN}{average Fourier norm}
\newabbreviation{maue}{MAUE}{mean absolute unbiased error}
\newabbreviation{mae}{MAE}{mean absolute error}
\newabbreviation{awgn}{AWGN}{additive white Gaussian noise}
\newabbreviation{std}{STD}{standard deviation}

\newabbreviation{ml}{ML}{maximum likelihood}
\newabbreviation{es}{ES}{exhaustive search}
\newabbreviation{bf}{BF}{brute force}
\newabbreviation{gd}{GD}{gradient descent}
\newabbreviation{gn}{GNA}{Gauss-Newton algorithm}
\newabbreviation{pinv}{PINV}{pseudo-inversion}
\newabbreviation{tsvd}{TSVD}{truncated \glsentrylong{svd}}
\newacronym{tsvdl}{TSVDL}{\glsentryshort{tsvd} with L-curve based parametric estimation}
\newabbreviation{gsvd}{GSVD}{generalized \glsentrylong{svd}}
\newabbreviation{tik}{RR}{ridge regression}
\newacronym{tikl}{RRL}{\glsentryshort{tik} with L-curve based parametric estimation}
\newabbreviation{wav}{WAV}{wavelet regularization}
\newabbreviation{ldct}{LDCT}{\glsentryshort{lasso}-\glsentryshort{dct}}
\newabbreviation{ldwt}{LDWT}{\glsentryshort{lasso}-\glsentryshort{dwt}}
\newabbreviation{ltv}{LTV}{\glsentryshort{lasso}-\glsentryshort{tv}}

\newabbreviation{gce}{GCE}{Gaussian-fit center estimation}
\newabbreviation{cce}{CCE}{centroid center estimation}
\newabbreviation{sce}{SCE}{scanline center estimation}
\newabbreviation{mle}{MLE}{maximum likelihood estimation}
\newabbreviation{rv}{r.v.}{random variable}
\newabbreviation{fft}{FFT}{fast Fourier transform}
\newabbreviation{dft}{DFT}{discrete Fourier transform}
\newabbreviation{dct}{DCT}{discrete cosine transform}
\newabbreviation{dwt}{DWT}{discrete wavelet transform}
\newabbreviation{swt}{SWT}{stationary wavelet transform}
\newabbreviation{qmf}{QMF}{quadrature mirror filter}
\newabbreviation{lpf}{LPF}{low pass filter}
\newabbreviation{fir}{FIR}{finite impulse response}
\newabbreviation{iir}{IIR}{infinite impulse response}
\newabbreviation{idct}{IDCT}{inverse \glsentrylong{dct}}
\newabbreviation{idwt}{IDWT}{inverse \glsentrylong{dwt}}
\newabbreviation{rip}{RIP}{restricted isometry property}
\newabbreviation{lasso}{LASSO}{least absolute shrinkage and selection operator}
\newabbreviation{omp}{OMP}{orthogonal matching pursuit}
\newabbreviation{admm}{ADMM}{alternating direction method of multipliers}

\newabbreviation{tv}{TV}{total variation}
\newabbreviation{rof}{ROF}{Rudin-Osher-Fatemi}
\newabbreviation{atv}{ATV}{anisotropic \glsentrylong{tv}}
\newabbreviation{itv}{ITV}{isotropic \glsentrylong{tv}}
\newabbreviation{vtv}{VTV}{vector \glsentrylong{tv}}
\newabbreviation{ctv}{CTV}{collaborative \glsentrylong{tv}}
\newabbreviation{stv}{STV}{Shannon \glsentrylong{tv}}
\newabbreviation{utv}{UTV}{upwind \glsentrylong{tv}}
\newabbreviation{tgv}{TGV}{total generalized variation}
\newabbreviation{als}{ALS}{alternating least squares}
\newabbreviation{mm}{MM}{majorization-minimization}

\newabbreviation{cfa}{CFA}{color filter array}
\newabbreviation{sfa}{SFA}{spectral filter array}
\newabbreviation{msfa}{MSFA}{\glsentrylong{ms} filter array}

\newabbreviation{cnn}{CNN}{convolutional neural network}
\newabbreviation{fts}{FTS}{Fourier transform spectrometer}
\newabbreviation{ftir}{FTIR}{Fourier transform infrared spectroscopy}
\newabbreviation{em}{EM}{electro-magnetic}
\newabbreviation{tem}{TEM}{transverse electro-magnetic}
\newabbreviation{te}{TE}{transverse electric}
\newabbreviation{tm}{TM}{transverse magnetic}
\newabbreviation{fsr}{FSR}{free spectral range}
\newabbreviation{fwhm}{FWHM}{full width at half maximum}
\newabbreviation{fpa}{FPA}{focal plane array}
\newabbreviation{fov}{FoV}{field of view}
\newabbreviation{aoa}{AoA}{angle of acceptance}
\newabbreviation{fp}{FP}{Fabry-Pérot}
\newabbreviation{opd}{OPD}{optical path difference}
\newabbreviation{opl}{OPL}{optical path length}
\newabbreviation{ccd}{CCD}{charged coupled device}
\newabbreviation{cmos}{CMOS}{complementary metal-oxide-semiconductor}
\newabbreviation{sar}{SAR}{synthetic aperture radar}
\newabbreviation{lidar}{LIDAR}{light detection and ranging}
\newabbreviation{uav}{UAV}{unmanned aerial vehicle}
\newacronym{csens}{CoSe}{compressed sensing}
\newabbreviation{cp}{CP}{compressed pansharpening}
\newabbreviation{ft}{FT}{Fourier transform}
\newabbreviation{psf}{PSF}{point spread function}
\newabbreviation{ffc}{FFC}{flat field correction}
\newabbreviation[longplural={regions of interest}]{roi}{RoI}{region of interest}

\newabbreviation[longplural={greenhouse gases}]{ghg}{GHG}{greenhouse gas}
\newabbreviation{ehs}{EHS}{environment, health and safety}
\newabbreviation{fpga}{FPGA}{field programmable gate array}

\newabbreviation{qb}{QB}{QuickBird}
\newabbreviation{wv2}{WV2}{WorldView-2}
\newabbreviation{wv3}{WV3}{WorldView-3}
\newabbreviation{iko}{IKO}{IKONOS}
\newabbreviation{ge1}{GE-1}{GeoEye-1}
\newabbreviation{prisma}{PRISMA}{Hyperspectral Precursor of the Application Mission}
\newabbreviation{jpeg}{JPEG}{Joint Photographic Experts Group}
\newabbreviation{jp2k}{JP2K}{\glsentryshort{jpeg} 2000}
\newabbreviation{ccsds}{CCSDS}{Consultative Committee for Space Data Systems}

\newabbreviation{fui}{FUI}{Fonds Unique Interministériel}
\newabbreviation{anr}{ANR}{Agence Nationale de Recherche}
\newabbreviation{ipag}{IPAG}{Institut de Planétologie et d'Astrophysique de Grenoble}
\newabbreviation{onera}{ONERA}{Office National d'Etudes et de Recherches Aérospatiales}
\newabbreviation{imspoc}{ImSPOC}{image spectrometer on chip}
\newabbreviation[type=ignored]{imagaz}{ImaGAZ}{ImaGAZ}
\newabbreviation{scarbo}{SCARBO}{Space CARBon Observatory}
\newabbreviation[type=ignored]{imspoc-uv}{ImSPOC-UV}{\glsentryshort{imspoc}-\glsentrylong{uv}}
\newabbreviation{presto}{PRESTO}{Precursory Research for Embryonic Science and Technology}
\newabbreviation{nasa}{NASA}{National Aeronautics and Space Administration}
\newabbreviation{h2020}{H2020}{Horizon 2020}
\newabbreviation{fumultispoc}{FuMultiSPOC}{FUsion MULTIspectral-\glsentryshort{imspoc}}

\newabbreviation{p1}{PROTO-1}{prototype \glsentryshort{imspoc-uv}/\glsentryshort{vis}}
\newabbreviation{p2}{PROTO-2}{prototype \glsentryshort{imspoc-uv}-drone}
\newabbreviation{p3}{PROTO-3}{prototype \glsentryshort{imagaz}-1}
\newabbreviation{p4}{PROTO-4}{prototype NanoCarb-1}

\newacronym{phd}{PhD}{doctor of philosophy}

\begin{document}
\title{Joint Demosaicing and Fusion of Multiresolution Coded Acquisitions: A Unified Image Formation and Reconstruction Method}

\author{
	Daniele Picone\,\orcidlink{0000-0002-0226-8399},~\IEEEmembership{Member,~IEEE,}
	Mauro Dalla Mura\,\orcidlink{0000-0002-9656-9087},~\IEEEmembership{Senior Member,~IEEE,}\\
	and Laurent Condat\,\orcidlink{0000-0001-7087-1002},~\IEEEmembership{Senior Member,~IEEE}

	\thanks{D. Picone is with
		Univ. Grenoble Alpes, CNRS, Inria, Grenoble INP, GIPSA-lab,
		38000 Grenoble, France.
		D. Picone is also with
		Univ. Grenoble Alpes, CNRS, Grenoble INP, IPAG,
		38000 Grenoble, France (e-mail: daniele.picone@grenoble-inp.fr).
	}
	\thanks{M. Dalla Mura is with Univ. Grenoble Alpes, CNRS, Inria, Grenoble INP, GIPSA-lab, 38000 Grenoble, France. M. Dalla Mura is also with Institut Universitaire de France (IUF), 75005 Paris, France (e-mail: mauro.dalla-mura@grenoble-inp.fr).}
	\thanks{L. Condat is with King Abdullah University of Science and Technology (KAUST), 23955 Thuwal, Saudi Arabia (e-mail: laurent.condat@kaust.edu.sa).
	}
	\thanks{This work is partly supported by grant ANR FuMultiSPOC (\href{https://anr.fr/Project-ANR-20-ASTR-0006}{ANR-20-ASTR-0006}).}
	\thanks{This paper has supplementary downloadable material available at \href{http://doi.org/10.1109/TCI.2023.3261503}{http://doi.org/10.1109/TCI.2023.3261503}, provided by the authors. \textit{(Corresponding author: Mauro Dalla Mura)}}
}

\markboth{Journal of \LaTeX\ Class Files,~Vol.~14, No.~8, August~2021}%
{Picone \MakeLowercase{\textit{et al.}}: Joint demosaicing and fusion of multiresolution coded acquisitions: {A} unified image formation and reconstruction method}

\fancypagestyle{FirstPage}{
	\fancyhf{}
	\renewcommand{\headrulewidth}{0pt}
	\renewcommand{\footrulewidth}{0pt}
	\renewcommand{\footskip}{20pt}
	\fancyfoot[C]{\footnotesize {© 2023 IEEE. Personal use of this material is permitted.  Permission from IEEE must be obtained for all other uses, in any current or future media, including reprinting/republishing this material for advertising or promotional purposes, creating new collective works, for resale or redistribution to\\servers or lists, or reuse of any copyrighted component of this work in other works.}} 
}

\maketitle

\begin{abstract}
Novel optical imaging devices allow for hybrid acquisition modalities such as compressed acquisitions with locally different spatial and spectral resolutions captured by a single focal plane array.
In this work, we propose to model the imaging camera system for a \gls{mrca} in a unified framework, which includes conventional devices such as those based on spectral/color filter arrays, compressed coded apertures, and multiresolution sensing. We also propose a model-based image reconstruction algorithm performing a \gls{jodefu} of any acquisition modeled in the \gls{mrca} framework. The \gls{jodefu} reconstruction algorithm solves an inverse problem with a proximal splitting technique and is able to reconstruct an uncompressed image datacube at the highest available spatial and spectral resolution. An implementation of the code is available at \href{https://github.com/danaroth83/jodefu}{https://github.com/danaroth83/jodefu}.
\end{abstract}

\glsresetall


\maketitle

\IEEEpeerreviewmaketitle

\section{Introduction}
\label{sec:intro}

\thispagestyle{FirstPage}

	\IEEEPARstart{C}{onventional} cameras acquire images that are immediately exploitable by the end user with little or no processing of the raw acquisition. When the acquisition relies on a spectral or spatial scanning of the scene, these cameras provide the image as a \textit{datacube} with spatial and spectral dimensions~\cite{Eism12, Mano16}.
	For example, this is the case of \gls{rgb}, \gls{ms} and \gls{hs} imaging systems, where the datacube has either three, up to a few tens, or more channels, respectively.

	A different acquisition approach, following the \textit{computational imaging} paradigm~\cite{Stor09, Mira08, Coss13, Mitr14}, is based on \textit{compressed acquisitions}. In some cases, this allows to perform acquisitions that are instantaneous (i.e., snapshot) or with a lower number of acquired samples with respect to a conventional full scanning of the datacube. We in particular focus our attention to \textit{coded acquisitions}, for which the pixel on the focal plane can be described as a particular linear combination of the samples of the original datacube; this is for example the case of acquisitions captured through a \gls{cfa}, or where the input image is masked through any coded aperture before combining on the sensing system.
	As coded acquisitions~\cite{Brad09, Ster16} are not captured in the end user desired domain, a computational phase is needed to retrieve a datacube that is intelligible to the final user.
	In this work, we refer to such acquisition techniques as \textbf{image formation} methods, and to the required processing algorithms that recover the desired datacube as \textbf{image reconstruction} methods.

\begin{figure}
	\captionsetup[subfigure]{justification=centering}
	\centering
		\begin{subfigure}[t]{\linewidth}
			\includegraphics[width=0.99\linewidth]{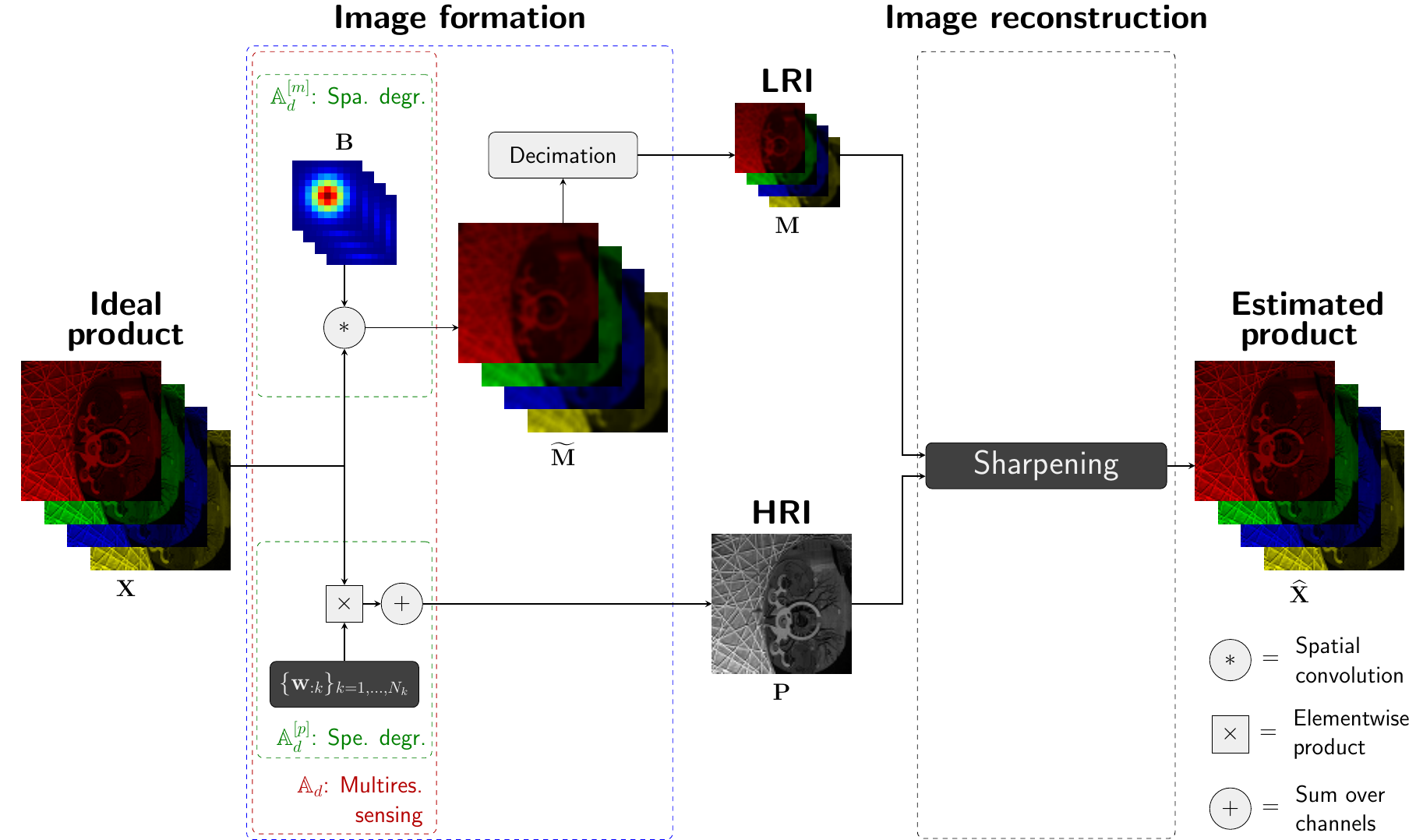}
			\caption{Degradation and sharpening}
			\label{fig:direct_classic_pansharpening}
		\end{subfigure}
		\smallskip
		
		\begin{subfigure}[t]{\linewidth}
			\includegraphics[width=0.99\linewidth]{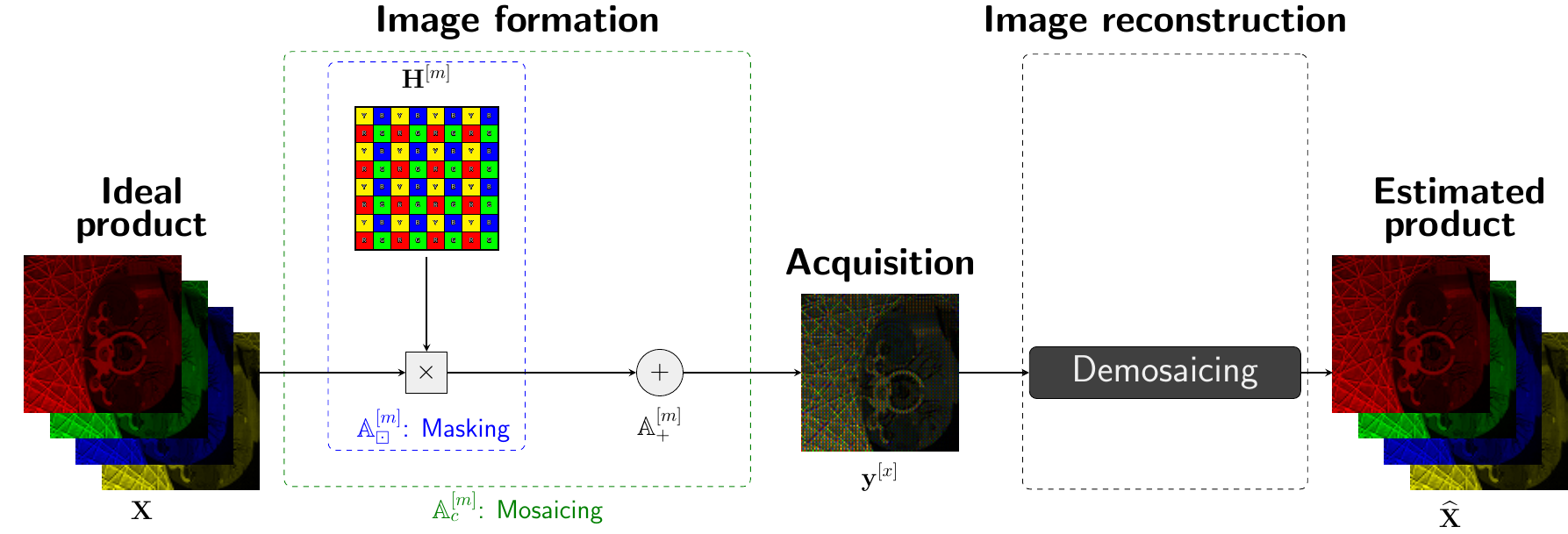}
			\caption{Mosaicing and demosaicing}
			\label{fig:direct_classic_demosaicing}
		\end{subfigure}
		\caption[Classic image formulation methods]{Classic formulation of the sharpening and demosaicing problem. The red, green, blue, and yellow slices represent each of the available channels of the image to reconstruct.}
		\label{fig:direct_classic}
	\end{figure}

	A wide selection of snapshot image formation methods are available in the literature~\cite{Yuan21, Zhou11}, which demands adequate techniques to process the acquisitions in a prompt and flexible manner. As the manufacture of many of such prototypes can be seen as a combination of elemental components, this work proposes to describe them under a unified mathematical reconstruction framework. In the proposed framework, the image formation step is described as a composition of basic operators which emulates the behavior of the elementary camera components. In particular, the work is focused on two classic acquisition scenarios, the multiresolution sensing and the mosaicing, which are described below:

	\begin{itemize}
		\item The \textbf{multiresolution sensing}, shown in \figurename\ref{fig:direct_classic_pansharpening}, is an acquisition setup where different sensor technologies are employed to provide complementary information of the same scene to be fused in the processing stage. This setup addresses the technical constraint of single sensors which are not capable of simultaneously achieving the spatial and spatial resolution desired by the final user.
		Most commonly, the product is available as a bundle of two images: the \gls{hri}, with high spatial and low spectral resolution, and the \gls{lri}, with low spatial and high spectral resolution. In the data fusion phase, known as \textit{sharpening}, the target is to produce a synthetic image with the highest available resolutions both in the spectral and spatial domain. This is a more generic formulation of the \textit{pansharpening} problem~\cite{Vivo15a, Lonc15}, where the target is to fuse a monochromatic acquisition, known as \gls{pan}, and a \glsentryfull{ms} image.
		\item The \textbf{mosaicing}, shown  in \figurename~\ref{fig:direct_classic_demosaicing}, is an acquisition technique where the captured samples are the output of a set of image sensors distributed over a \gls{fpa} and overlaid with an array of filters, known as \textit{\gls{cfa}} or \textit{\gls{msfa}}~\cite{Lapr14}. As an effect of filtering, each captured image pixel is associated to a given color/channel component and the full raw acquisition is composed by a mosaic of such components. For example, the reader may be familiar with \gls{cfa} designs such as the Bayer pattern~\cite{Baye76}, where the filters are arranged in periodic $2 \times 2$ \gls{rgb} squares with 2 repeated green filters placed on the opposite vertices. The associated image reconstruction method, known as \textit{demosaicing}, consists in recovering the full spectral component of the image at each available position in the \gls{fpa}.
	\end{itemize}

	In more sophisticated mosaicing-based compressed acquisition systems, a given pixel can also be expressed as a \textit{coding} and multiplexing of a portion of the datacube; in other words, each pixel represents an encoding of the input signal as it captures a generic linear combination of samples associated to different channels. For example, this is the case of the \textit{\gls{cassi}}~\cite{Arce14}, for which each available channel is firstly masked with a \glsentrylong{dmd} and shifted horizontally over the focal plane before recombination. More advanced variants of this original design have also been proposed, i.e. by introducing colored coded apertures in place of the traditional blocking/unblocking mechanisms~\cite{Arce14} or by combining the spectral mosaicing with the shearing in the sensing mechanism~\cite{Corr15}.

    For such acquisitions, many authors have shown that coding optimization is an important step to provide quality acquisitions.
    Specifically, the effectiveness of the encoding is often measured through associated metrics, and among those, the \textit{\glsentrylong{rip}} has proven to be a common choice to provide theoretical limits of the quality of the reconstruction.
    For the \gls{cassi}, this analysis has been performed in~\cite{Argu12} and later reinterpreted in~\cite{Para17} in terms of the coherence of the sensing matrix. While we consider this topic out of scope in this work, the interested reader may refer to the vast literature on the field for further information~\cite{Jala19, Cand08}.

	More recently, both the scientific community and device manufacturers are showing interest for the design of hybrid systems, where compressed acquisitions might have different spatial/spectral resolutions. Specifically, for image formation methods, a series of \textit{\glsentryshort{rgbw} patterns} were proposed where a set of wideband pixels are interleaved to a more classic \gls{rgb} pattern. This is the case, for example, of the \textit{Onyx} device by Teledyne e2v~\cite{web_Onyx} and some patents deposed by Kodak~\cite{Kwan17}. A similar effect could be obtained with the \textit{COLOR SHADES}~\cite{web_Shad}, a technology which allows for a fully customizable spectral response for each filter on the \gls{fpa}. We can interpret these acquisition as hybrid, since the wideband pixels can be seen as a binning in the spectral domain, and consequently have a different spectral resolution with respect to the color ones.

	Modern commercial \gls{cfa} patterns, such as the \textit{Quad Bayer}, are also starting to implement mechanism of spatial binning across adjacent pixels, in order to improve the \glsentryshort{snr} of the detected photons in conditions of low illumination~\cite{Agra17}. More recently, a novel technology was proposed to focus the incident light rays over photodiodes through customly manufactured microlenses~\cite{Lakc18}. As a consequence, the resulting samples are effectively at a lower spatial resolution with respect to non-binned or less focused alternatives.

	Espitia et al.~\cite{Espi16} suggested to capture both the \gls{lri} and the \gls{hri} as separate \gls{cassi} acquisitions, introducing a joint Bayesian framework for the reconstruction of the full resolution image. An optical testbed of manufacturing feasibility for such device is reported in~\cite{Rued21}. Fu et al. \cite{Fu18} proposed instead a reconstruction algorithm where a \gls{hs} image is fused with a \gls{ms} mosaiced image obtained with a \gls{cfa} acquisition system. Takeyama and Ono~\cite{Take20} address the \textit{\glsentrylong{cp}}, where the quality of a noisy \gls{hri} is restored with the help of an associated compressed acquisition of a \gls{lri}.

	Additionally, such hybrid devices are a reasonable proof that, in the near future, multiresolution sensors could potentially be accommodated over the same focal plane.
    Such unconventional coded instruments can bring benefits in terms of reduced size, cost, and acquisition time with respect to traditional hyperspectral camera. In the field of remote sensing, most high-end commercial satellites are equipped with scanners (e.g., pushbroom).
    However, commercial off-the-shelf components such as consumer cameras based on masking are often used with drones or smallsats (e.g., AmicalSat~\cite{web_amicalsat}), and make up for an ideal application field for our work.

	Armed with this knowledge, the main aim of this work is to develop a unified framework for both the image formation and its reconstruction which includes all the previously cited examples. For this reason, in the context of image formation methods, we propose the \textbf{\gls{mrca}} framework, a formalization of the multiresolution acquisition model with compressed acquisitions.
	This model can formalize the design of an optical device based on the assumption that sensors with different characteristics can be accommodated on the same \gls{fpa} and that the resulting compressed acquisition contains partial information both from the \gls{lri} and the \gls{hri}, whose physical implementation was presented in our previous work~\cite{Pico18}.
	An example of the acquisitions of such device is shown in \figurename~\ref{fig:direct}, where the \gls{mrca} models the acquisition of a monochromatic raw image which includes samples from both a \gls{pan} and a \gls{ms}, according to arrangement shown in \figurename~\ref{fig:cfa_pattern_ubt4}.

	Since we have a unified acquisition model, we also propose the \textbf{\gls{jodefu}}, a generic image reconstruction algorithm that addresses both the demosaicing and the multiresolution fusion, extending the definition from our previous work~\cite{Pico18a}. The proposed algorithm both recovers the missing information of the compressed acquisition and fuses the multiresolution samples to reach the maximum available spatial and spectral resolution. The algorithm, which makes use of a Bayesian framework, is not exclusively a demosaicing-style image reconstruction, as we aim to reconstruct a fused product, nor it is a simple fusion, as the acquisition is not given by  well-distinguished multimodal sources, but rather by a lossy compressed combination of the two.

    In short, the novel contributions of this work include:
    \begin{itemize}
        \item the definition of the \gls{mrca}, a flexible model for multiresolution sensors sharing a common focal plane, which includes a series of well-known image formation methods, such as the \gls{cfa}~\cite{Naka05_short} acquisitions and the multiresolution sensing;
        \item the derivation of some properties of the direct model operator associated to the \gls{mrca}, which enable its use with proximal algorithms;
        \item the definition of the \gls{jodefu}, an image reconstruction framework capable of simultaneously addressing the problem of demosaicing and the fusion of partial multiresolution acquisition;
        \item a comparison of the performances of the \gls{jodefu} with respect to classic image reconstruction methods for compressed acquisitions; we also analyze the reconstructed products taken with the proposed \gls{mrca}, when they embed a different amount of \gls{ms} bands, and compare the \gls{jodefu} with alternative reconstruction approaches obtained by cascading a set of classical algorithms.
    \end{itemize}

    The paper is organized as follows: Section~\ref{sec:notation} introduces the notation, Section~\ref{sec:formation} describes the \gls{mrca}, highlighting how it expands on classic image formation methods; Section~\ref{sec:reconstruction} presents the \gls{jodefu} algorithm, and Section~\ref{sec:experiments} provides the related experiments.

    \section{Notation}
    \label{sec:notation}

    In this paper, we denote:
    \begin{itemize}
    	\item \textit{scalars} with lowercase non-bold letters (e.g. $u$);
    	\item \textit{vectors} with lowercase bold letters (e.g. $\mathbf{u}$);
    	\item \textit{matrices} with uppercase bold letters (e.g. $\mathbf{U}$);
    	\item \textit{tensors} (that is, arrays with more than two dimensions) with bold italic fonts (e.g. $\tens{U}$).
    \end{itemize}
	A detailed description of the variables used in this paper is given in~\tablename~\ref{tab:notation}. In this work, the image samples are either organized in their classic form, as a 3-way tensor whose dimensions represent the rows, columns and channels (e.g., $\tens{U}^{[x]}\in\mathbb{R}^{N_i \times N_j \times N_k}$), or in \textit{lexicographic order}, where the first two dimensions of the classic form are concatenated into one. As both representations contain the same samples, we can switch from the first to the second form without any loss of information, and we denote this operation with $\lexi(\cdot)$ (e.g., $\mathbf{X}=\lexi\left(\tens{U}^{[x]}\right)$, where $\mathbf{X}\in\mathbb{R}^{N_i N_j \times N_k}$).

    When we select a generic $k$-th slice of the image, this is denoted with a subscribed index $k$, while the non-sliced dimensions are denoted with a colon (e.g., $\mathbf{U}^{[x]}_{::k}$ denotes the $k$-th band of $\tens{U}^{[x]}$). When confusion may arise, the subscribed indices are separated by a comma (e.g., $\mathbf{U}^{[x]}_{:,:,k}$ is equivalent to  $\mathbf{U}^{[x]}_{::k}$). Once again, a more detailed description of such operations is shown in~\tablename~\ref{tab:notation}.

    Finally, $\|\cdot\|_2$ and $\|\cdot\|_F$ denote the $\ell_2$ and Frobenius norm, respectively.

    \begin{table*}
	\centering
	\caption[Notation]{Notation for the multidimensional arrays defined in this paper. Each variable can be expressed either in their classical order or by its lexicographic order, by reshaping the spatial dimensions of the former into a single dimension.}
	
	\begin{adjustbox}{width=0.9\linewidth}
		\begin{tabular}{l|ccccc|ccccc}
			\hline
			&\multicolumn{5}{c|}{\textbf{Classic representation}}&\multicolumn{5}{c}{\textbf{Lexicographic order representation}}\\
			\hline
			\multirow{2}{*}{\textbf{Variable}}&\multirow{2}{*}{\textbf{Symbol}}&\textbf{Dimensions}&$k$\textbf{-th}&$(i,j)$\textbf{-th}&$(i, j,k)$\textbf{-th}&\multirow{2}{*}{\textbf{Symbol}}&\textbf{Dimensions}&$k$\textbf{-th}&$i$\textbf{-th}&$(i,k)$\textbf{-th}\\
			&&\textbf{(Row} $\times$ \textbf{col.} $\times$ \textbf{band)}&\textbf{band}&\textbf{pixel}&\textbf{element}&&\textbf{(Pixel} $\times$ \textbf{band)}&\textbf{band}&\textbf{pixel}&\textbf{element}\\
			
			\hline
			\textbf{Reference}&$\tens{U}^{[x]}$&$N_i \times N_j \times N_k$&$\mathbf{U}^{[x]}_{::k}$&$\mathbf{u}^{[x]}_{ij:}$&$u^{[x]}_{ijk}$&$\mathbf{X}$&$N_i N_j \times N_k$&$\mathbf{x}_{:k}$&$\mathbf{x}_{i:}$&$x_{ik}$\\
			\textbf{\glsentryshort{hri}}&$\tens{U}^{[p]}$&$N_i \times N_j \times N_p$&$\mathbf{U}^{[p]}_{::k}$&$\mathbf{u}^{[p]}_{ij:}$&$u^{[p]}_{ijk}$&$\mathbf{P}$&$N_i N_j \times N_k$&$\mathbf{p}_{:k}$&$\mathbf{p}_{i:}$&$p_{ik}$\\
			\textbf{\glsentryshort{lri}}&$\tens{U}^{[m]}$&$\frac{N_i}{\ratio} \times \frac{N_j}{\ratio} \times N_k$&$\mathbf{U}^{[m]}_{::k}$&$\mathbf{u}^{[m]}_{ijk}$&$u^{[m]}_{ijk}$&$\mathbf{M}$&$\frac{N_i N_j}{\ratio^2} \times N_k$&$\mathbf{m}_{:k}$&$\mathbf{m}_{i:}$&$m_{ik}$\\
			\textbf{Upscaled \glsentryshort{lri}}&$\tens{U}^{[\widetilde{m}]}$&$N_i \times N_j \times N_k$&$\mathbf{U}^{[\widetilde{m}]}_{::k}$&$\mathbf{u}^{[\widetilde{m}]}_{ij:}$&$u^{[\widetilde{m}]}_{ijk}$&$\widetilde{\mathbf{M}}$&$N_i N_j \times N_k$&$\widetilde{\mathbf{m}}_{:k}$&$\widetilde{\mathbf{m}}_{i:}$&$\widetilde{m}_{ik}$\\
			\textbf{Acquisition}&$\mathbf{Y}$&$N_i \times N_j \times 1$&-&$y_{ij}$&$y_{ij}$&$\mathbf{y}$&$N_i N_j \times 1$&-&$y_i$&$y_i$\\
			\textbf{Estimated product}&$\tens{U}^{[\widehat{x}]}$&$N_i \times N_j \times N_k $&$\mathbf{U}^{[\widehat{x}]}_{::k}$&$\mathbf{u}^{[\widehat{x}]}_{ij:}$&$u^{[\widehat{x}]}_{ijk}$&$\widehat{\mathbf{X}}$&$N_i N_j \times N_k$&$\widehat{\mathbf{x}}_{:k}$&$\widehat{\mathbf{x}}_{i:}$&$\widehat{x}_{ik}$\\
			\textbf{Blurring kernels}&$\tens{U}^{[b]}$&$N_b \times N_b \times N_k$&$\mathbf{U}^{[b]}_{::k}$&$\mathbf{u}^{[b]}_{ij:}$&$u^{[b]}_{ijk}$&$\mathbf{B}$&$N_b^2 \times N_k$&$\mathbf{b}_{:k}$&$\mathbf{b}_{i:}$&$b_{ik}$\\
			\textbf{\glsentryshort{hri} mask}&-&-&-&-&-&$\mathbf{H}^{[p]}$&$N_i N_j \times N_p$&$\mathbf{h}^{[p]}_{:k}$&$\mathbf{h}^{[p]}_{i:}$&$h^{[p]}_{ik}$\\
			\textbf{\glsentryshort{lri} mask}&-&-&-&-&-&$\mathbf{H}^{[m]}$&$N_i N_j \times N_k$&$\mathbf{h}^{[m]}_{:k}$&$\mathbf{h}^{[m]}_{i:}$&$h^{[m]}_{ik}$\\
			\hline
		\end{tabular}
	\end{adjustbox}
	\label{tab:notation}
\end{table*}

\section{Proposed image formation model}
\label{sec:formation}

   	In this section, we introduce the mathematical model of the \gls{mrca} (Section~\ref{ssec:formation_joint}) and its properties (Section~\ref{ssec:formation_properties}). As a preliminary step, we firstly define the two main component of our image formation models: the multiresolution sensing in Section~\ref{ssec:formation_degradation} and the mosaicing  in Section~\ref{ssec:formation_mosaicing}.

   	\subsection{Multiresolution sensing}
   	\label{ssec:formation_degradation}

   		When multiresolution acquisitions are involved, each of the sensor technologies is characterized by a limited spatial and spectral resolution.
   		Therefore, in the most general case, the multiresolution sensing setup is expressed as a set of multiple acquisitions, which present a certain spatial and/or spectral degradation with respect to the ideal datacube to reconstruct $\mathbf{X}$.
   		For the sake of exposition but without loss of generalization, we limit our analysis to the most common scenario in the literature~\cite{Vivo15a,Lonc15,Alpa15}, in which the acquisition is composed by an \gls{hri} and an \gls{lri}. The \gls{hri} $\mathbf{P}$ and the \gls{lri} $\widetilde{\mathbf{M}}$ are respectively obtained as a degradation in the spectral and spatial domain, so that:
   		\begin{equation}
   			\begin{cases}
   				\mathbf{P}&=\lino{A}\nolimits_d^{[p]}(\mathbf{X})\,,\\
   				\widetilde{\mathbf{M}}&=\lino{A}\nolimits_d^{[m]}(\mathbf{X})\,.
   			\end{cases}
   			\label{eq:direct_multiresolution}
   		\end{equation}
   		In the previous equation:
   		\begin{itemize}
   			\item the \textbf{spectral degradation} operation $\mathbf{P}=\lino{A}_d^{[p]}(\mathbf{X})$ is described by a linear combination in the form:
   			\begin{equation}
   				\mathbf{p}_{:j}=\sum\limits_{k=1}^{N_k} w_{jk} \mathbf{x}_{:k}\,,\;\;\;\;\;\;\forall \range{j}{1}{N_p}
   				\label{eq:spectral_degradation}
   			\end{equation}
   			where $\{w_{jk}\}_{\range{j}{1}{N_p},\,\range{k}{1}{N_k}}$ are the \textit{weight coefficients} associated to the spectral responses of the sensors.
   			This is a widespread choice in the literature, as the sensors perform an integration of the incoming radiance that is modulated by the spectral response of the filters~\cite{Eism12}.
   			\item the \textbf{spatial degradation} operation $\widetilde{\mathbf{M}}=\lexi(\tens{U}^{[\widetilde{m}]})=\lino{A}_d^{[m]}(\mathbf{X})$ is a convolution by a set of filters  $\tens{U}^{[b]}\in\mathbb{R}^{N_b \times N_b \times N_k}$. Specifically:
   			\begin{equation}
   				\mathbf{U}^{[\widetilde{m}]}_{::k} = \mathbf{U}^{[x]}_{::k} \cnvo \mathbf{U}^{[b]}_{::k}\,,\;\;\;\;\;\;\forall \range{k}{1}{N_k}\,,
   				\label{eq:spatial_degradation}
   			\end{equation}
   			where $\cnvo$ denotes a spatial convolution operator.
   			Once again, this is a widely employed model in the literature, derived from the assumption that the blurring effect of the sensors can be described by a linear translation-invariant operator, and hence uniquely defined by its \textit{\gls{mtf}}~\cite{Alpa15}.
   		\end{itemize}

   		Finally, a decimation of $\widetilde{\mathbf{M}}$  by a scale factor $\ratio$ (i.e., taking every $\ratio$ samples both in the vertical and horizontal direction) produces the actual \gls{lri} $\mathbf{M}$.

   	\subsection{Mosaicing}
   	\label{ssec:formation_mosaicing}

    	In the classic formulation~\cite{Lapr14} shown in \figurename~\ref{fig:direct_classic_demosaicing}, the \gls{cfa}/\gls{msfa}-based mosaicing is modeled as an element-wise multiplication of the input by a mask $\mathbf{H}\in\mathbb{R}^{N_i N_j \times N_k}$.
    	Among those, \textit{binary masks} define a special case of $\mathbf{H}$ whose elements can only be either zeros or ones; in other terms, each pixel of the mask identifies a specific channel, which is transferred on the focal plane. This allows them to be represented by a color-coded matrix, such as those shown in \figurename~\ref{fig:cfa_pattern}.

    	In this work, we model the acquisition $\mathbf{y}=\lino{A}\nolimits_c^{[m]}(\mathbf{X})$ due to the mosaicing as the cascade of the following two operations:
    	\begin{itemize}
    		\item \textbf{Masking} $\mathbf{X}\masked=\lino{A}_\boxdot^{[m]}(\mathbf{X})$: where an element-wise multiplication (denoted by $\hadam$) is applied independently on each band, yielding:
	    	\begin{equation}
	    		\mathbf{x}\masked_{:k} = \mathbf{x}_{:k} \hadam \mathbf{h}_{:k}\,,\;\;\;\;\;\;\forall \range{k}{1}{N_k}\,.
	    		\label{eq:masking}
	    	\end{equation}
    		If the mask is binary, the variable $\mathbf{x}\masked_{:k}$ is commonly known as \textbf{sparse channel} in the demosaicing literature~\cite{Miho17}, since the element-wise multiplication sets most of its values to zero.
    		\item \textbf{Sum over channels} $\mathbf{y}=\lino{A}^{[m]}_+(\mathbf{X}\masked)$: where the final observation is obtained by summing along the spectral dimension, returning:
    		\begin{equation}
    			\mathbf{y}= \sum\limits_{k=1}^{N_k} \mathbf{x}\masked_{:k} = \sum\limits_{k=1}^{N_k}\mathbf{x}_{:k} \hadam \mathbf{h}_{:k}\,.
    			\label{eq:sum_over_channels}
    		\end{equation}
    	\end{itemize}

    	In some more advanced acquisition devices, some optical elements allow to shift the captured light rays of a given channel over the \gls{fpa}.
    	To model such effect, we introduce an additional \textbf{shifting operator} $\lino{A}_\rightarrow^{[m]}$ in between the previously presented ones, so that the full operation of mosaicing becomes:
    	\begin{equation}
    		\mathbf{y}=\lino{A}\nolimits_c^{[m]}(\mathbf{X})=\lino{A}\nolimits_+^{[m]}\left(\lino{A}\nolimits_\rightarrow^{[m]}\left(\lino{A}\nolimits_\boxdot^{[m]}(\mathbf{X})\right)\right)\,.
    		\label{eq:direct_mosaic}
    	\end{equation}

		The generic element $s_{ijk}$ of the shifted image $\tens{S}$ in the operation $\lexi(\tens{S})=\lino{A}_\rightarrow^{[m]}(\mathbf{X})$ is given by:
		\begin{equation}
			s_{ijk} = u^{[x]}_{\mathbf{r}(i,j,k)}\,,
			\label{eq:shift}
		\end{equation}
		where $\mathbf{r}(i,j,k):\mathbb{N}^3\rightarrow\mathbb{N}^3$ is a one-to-one vector function which defines the transformation from a source to a target position.

		For example, in the case of the single dispersion \gls{cassi}~\cite{Arce14}, the focal plane associated to each channel
		can be rigidly translated in the horizontal direction through a diffraction prism.
    	In the problem defined by the original authors, this is a shift by one pixel between adjacent channels, which we reformulated within the \gls{mrca} framework by defining a shifted input $\tens{S}\in\mathbb{R}^{N_i \times (N_j+N_k-1) \times N_k}$ such that:
    	\begin{equation}
    		\mathbf{r}(i,j,k) = (i, j+k-1, k)\,.
    		\label{eq:cassi}
    	\end{equation}

	\subsection{\texorpdfstring{\glsreset{mrca}\Gls{mrca}}{Multiresolution compressed acquisition (MRCA)}}
	\label{ssec:formation_joint}

\begin{figure*}
    \centering
    \includegraphics[width=0.95\linewidth]{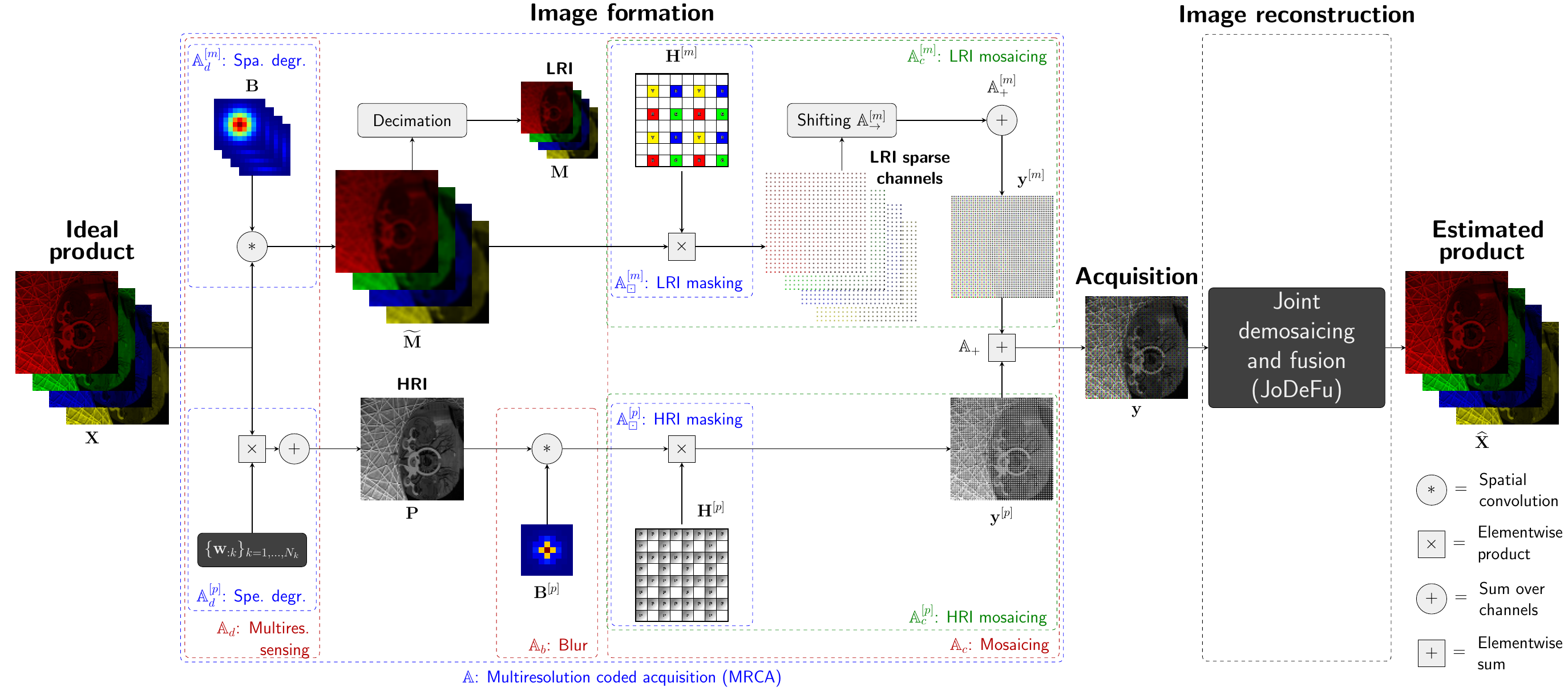}
    \caption[Image formation of the \glsentryshort{mrca}]{
    Representative scheme of the image formation model for the proposed \gls{mrca} framework. The multiresolution configuration is represented by a \gls{pan} and by a color coded 4-band \gls{ms}; the white spots in the masks denote zero pixels in all bands. The \gls{hri} shifting $\lino{A}^{[p]}_\rightarrow$ and summing $\lino{A}^{[p]}_+$ operators are not depicted in the figure.}
    \label{fig:direct}
\end{figure*}

    	The main target of the proposed \gls{mrca} is to allow for multiresolution images to be stored over the same focal plane. The proposed model $\mathbf{y}=\lino{A}(\mathbf{X})$ for the acquisition system is shown in \figurename~\ref{fig:direct}, and is expressed as:
    	\begin{equation}
    		\lino{A}(\mathbf{X})=\lino{A}\nolimits_+^{ }\left(\lino{A}\nolimits_b^{ }(\lino{A}\nolimits_c^{[p]}(\lino{A}\nolimits_d^{[p]}\left(\mathbf{X}))\right),\;\; \lino{A}\nolimits_c^{[m]}(\lino{A}\nolimits_d^{[m]}(\mathbf{X}))\right)\,.
    		\label{eq:mrca}
    	\end{equation}

    	The operator $\lino{A}$ is given by the following cascade of operations:
    	\begin{itemize}
    		\item A \textbf{multiresolution sensing operator} $\lino{A}_d$, composed by the operators $\lino{A}_d^{[m]}$ from eq.~\eqref{eq:spatial_degradation} and $\lino{A}_d^{[p]}$ from eq.~\eqref{eq:spectral_degradation} of Section~\ref{ssec:formation_degradation}, which generate the \gls{lri} and \gls{hri} branch, respectively;
    		\item A \textbf{mosaicing operator} $\lino{A}_c$: where the operator $\lino{A}_c^{[m]}$ of eq.\eqref{eq:direct_mosaic} is applied on the \gls{lri} branch and $\lino{A}_c^{[p]}$, identical to $\lino{A}_c^{[m]}$ except for acting over $N_p$ channels instead of $N_k$, and applied to the \gls{hri} branch.
    		\item A \textbf{blur operator} $\lino{A}_b$, to adjust its scale ratio of the \gls{hri} with respect to the reference and modeled as a spatial convolution.
    		\item A \textbf{sum operator} $\lino{A}_+(\mathbf{y}^{[m]},\; \mathbf{y}^{[p]})$, which sums the \gls{lri} mosaic $\mathbf{y}^{[m]}$ and \gls{hri} mosaic $\mathbf{y}^{[m]}$ over the same focal plane:
    		\begin{equation}
    			\lino{A}\nolimits_+\left(\mathbf{y}^{[m]}, \mathbf{y}^{[p]}\right) = \mathbf{y}^{[m]} + \mathbf{y}^{[p]}\,.
    		\end{equation}
    	\end{itemize}

   		The reader may have noticed that the upscaled \gls{lri} $\widetilde{\mathbf{M}}$ is not decimated in the \gls{mrca} pipeline; this step is unnecessary in our framework as the mask $\mathbf{H}^{[m]}$ can automatically suppress the pixel values that would be eliminated during the decimation process.

	    The \gls{mrca} can model (assuming that no blur is introduced by $\lino{A}_b$):
	    \begin{itemize}
   	    	\item \textit{multiresolution sensing}, if both the operators $\lino{A}_c$ and $\lino{A}_+$ are substituted with an identity, as eq.~\eqref{eq:mrca} reduces to eq.~\eqref{eq:direct_multiresolution};
   	    	\item \textit{mosaicing}, if both operators in $A_d^{[m]}$ is an identity and $\mathbf{y}^{[p]}$ is an all-zero matrix, for which eq.~\eqref{eq:mrca} reduces instead to eq.~\eqref{eq:direct_mosaic}.
	    \end{itemize}
    	Therefore, the \gls{mrca} framework is a general formation method for acquisition systems that involves multiresolution sensing and mosaicing.
    	A more detailed description of the operators for such special cases, which also includes the frameworks proposed by Li et al.~\cite{Li19} and by Takeyama and Ono~\cite{Take20}, is given in~\tablename~\ref{tab:mrca}.
    	Additionally, the framework is also capable to model some more advanced sensing mechanism inspired by the CASSI, such as the one with colored coded apertures proposed in~\cite{Argu14}. For such case, as the spectral bands are combined according to the spectral response of each pixel of the coded aperture, the coefficients of the mask are non binary. By inverting the shearing and mosaicing operators, the design can also be extended to the colored CASSI~\cite{Corr15}.

	    \begin{table}
	\newcolumntype{A}{m{10pt}}
	\newcolumntype{B}{m{36pt}}
	
	\centering
	\caption[Image formation frameworks]{Description of classical image formation methods under the proposed \gls{mrca} framework.}
	
	\begin{adjustbox}{width=\linewidth}
	\begin{threeparttable}
		\begin{tabular}{l|AA|AAA|AAA|A}
			\hline
			&\multicolumn{2}{c|}{Multires.}&\multicolumn{3}{c|}{\glsentryshort{hri} mosaicing}&\multicolumn{3}{c|}{\glsentryshort{lri} mosaicing}&~\\
			\hline	
			\textbf{Image formation method}&$\lino{A}_d^{[p]}$&$\lino{A}_d^{[m]}$&$\lino{A}^{[p]}_\boxdot$&$\lino{A}_\rightarrow^{[p]}$&$\lino{A}_+^{[p]}$&$\lino{A}^{[m]}_\boxdot$&$\lino{A}_\rightarrow^{[m]}$&$\lino{A}_+^{[m]}$&$\lino{A}_+$\\\hline
			\glsentryshort{mrca}&\cmark&\cmark&\cmark&\cmark&\cmark&\cmark&\cmark&\cmark&\cmark\\
			Multiresolution sensing&\cmark&\cmark&\xmark&\xmark&\xmark&\cmark\tnote{1}&\xmark&\xmark&\xmark\\
			\Glsentryshort{cfa} mosaicing&\cmark\tnote{2}&\xmark&\xmark&\xmark&\cmark&\cmark&\xmark&\cmark&\cmark\\
			\Glsentryshort{cassi} acquisition~\cite{Arce14}&\cmark\tnote{2}&\xmark&\xmark&\xmark&\cmark&\cmark&\cmark\tnote{3}&\cmark&\cmark\\
			Lu et al.~\cite{Lu09}&\cmark&\cmark&\cmark&\xmark&\cmark&\xmark&\xmark&\xmark&\xmark\\
			Takeyama and Ono~\cite{Take20}&\cmark&\cmark&\xmark&\xmark&\xmark&\multicolumn{3}{c|}{\rule[2pt]{0.07\linewidth}{0.4pt}\hfil\cmark\tnote{4}     \hfil\rule[2pt]{0.07\linewidth}{0.4pt}}&\xmark\\
			\hline
		\end{tabular}
		\begin{tablenotes}\footnotesize
			\item[1] where $\mathbf{H}^{[m]}$ is a binary mask, with zeros at the interleaved positions of the \gls{lri};
			\item[2] with $w_{kl}=0\,,\, \forall\, \range{k}{1}{N_p},\,\range{l}{1}{N_k}$ to suppress \gls{hri} samples;
			\item[3] with the condition of eq.~\eqref{eq:cassi};
			\item[4] with any kind of linear compression is allowed for the \gls{lri}.\\
			In this table, \xmark~marks linear operators of the \gls{mrca} being substituted by an identity.
		\end{tablenotes}
	\end{threeparttable}
	\end{adjustbox}
	\label{tab:mrca}
\end{table}

   	 \subsection{Properties of the direct model operators}
   	 \label{ssec:formation_properties}

   	 	We investigate here two key properties of the operator $\lino{A}$, the adjoint operator and the operator norm, that are necessary to define the generic image reconstruction algorithm to be presented in Section~\ref{sec:reconstruction}.
   	 	Specifically:
   	 	\begin{itemize}
   	 		\item the \textbf{adjoint operator} $\lino{A}^*$ is the one that verifies the condition:
   	 		\begin{equation}
   	 			\left\langle\lino{A}(\mathbf{X}),\;\mathbf{y}\right\rangle=\left\langle\mathbf{X},\;\lino{A}\nolimits^*(\mathbf{y})\right\rangle\,,
   	 		\end{equation}
   	 		for all $\mathbf{X}\in\mathbb{R}^{N_iN_j \times N_k}$ and $\mathbf{y} \in\mathbb{R}^{N_iN_j}$. Here, $\langle\cdot,\;\cdot\rangle$ on the left and right side of the equation are the scalar products of the spaces of $\mathbf{y}$ and $\mathbf{X}$, respectively.
   	 		If $\lino{A}$ defines a matrix multiplication applied over vectorized inputs, $\lino{A}^*$ is equivalent of applying the Hermitian of the same matrix;
   	 		\item the \textbf{operator norm} $\|\lino{A}\|_{op}$ is defined~\cite{Rudi91} as the smallest scalar $\gamma$ such that:
   			\begin{equation}
				\|\lino{A}(\mathbf{X})\|_2\le \gamma\|\mathbf{X}\|_F\,,\;\;\;\;\;\;\forall\,\mathbf{X}\in\mathbb{R}^{N_iN_j \times N_k}\,.
				\label{eq:adjoint_definition}
			\end{equation}
			If $\lino{A}$ defines a matrix multiplication over vectorized inputs, its operator norm is the largest singular value of the matrix itself~\cite{Rudi91}.
   	 	\end{itemize}

   	 	For composed operators, their properties can be dynamically obtained by combining the elementary building blocks associated to the simple operations defined in the previous sections, in order to produce the image formation model under test.

    	Specifically, the adjoint of a composed operation $\lino{A}(\mathbf{X})=\lino{A}_c(\lino{A}_d(\mathbf{X}))$ is equal to applying the individual adjoint operators in reverse order:
    	\begin{equation}
    		\lino{A}\nolimits^*(\mathbf{y})=\lino{A}\nolimits^*_d(\lino{A}\nolimits^*_c(\mathbf{y}))\;.
    		\label{eq:adjoint_cascade}
    	\end{equation}
     	The only requirement to evaluate $\lino{A}^*$ then simplifies to deriving the adjoint operator of each elementary component of the \gls{mrca} separately.
    	These components can be split into the following categories:
    	\begin{itemize}
    		\item \textbf{Spatial convolution} (Operators $\lino{A}_d^{[m]}$, and $\lino{A}_b$): the adjoint of a convolution by a given kernel is a correlation by the same kernel;
    		\item \textbf{Sum over channels} (Operators $\lino{A}^{[m]}_+$, $\lino{A}^{[p]}_+$, and  $\lino{A}_+$): the adjoint is equivalent to replicating a monochromatic image across all bands;
    		\item \textbf{Shifting} (Operators $\lino{A}_{\rightarrow}^{[m]}$, and $\lino{A}_{\rightarrow}^{[p]}$): the adjoint of shifting a sample to a new position is a shift back to its original one.
    		\item \textbf{Element-wise product} (Operators $\lino{A}_d^{[p]}$, $\lino{A}_\boxdot^{[p]}$, and $\lino{A}_\boxdot^{[m]}$): the adjoint operator is itself, as this operation is self-adjoint.
    	\end{itemize}

   		We follow a similar approach for the operator norm. Specifically, we apply the Cauchy inequality to a composed operator $\lino{A}(\mathbf{x})=\lino{A}_c(\lino{A}_d(\mathbf{X}))$ to identify an upper limit for $\|\lino{A}\|_{op}$:
   		\begin{equation}
   			\|\lino{A}\|_{op}\le\|\lino{A}\nolimits_c\|_{op}\|\lino{A}\nolimits_d\|_{op}\,.
   			\label{eq:norm_cascade}
   		\end{equation}
   		This inequality can be substituted with a strict equality as the reconstruction algorithms we employ only require upper bounds for the operator norm. Nonetheless, we can once again separate the problem into evaluating the operator norms individually:
    	\begin{itemize}
			\item \textbf{Spatial convolution of} $\mathbf{X}$ \textbf{by} $\mathbf{B}$: for each band, the convolution by the $k$-th band $\mathbf{b}_{:k}$ can be rewritten as a multiplication by a circulant matrix. Its singular values are hence defined as the sum of the coefficients $\mathbf{b}_{:k}$ weighted by the complex roots of unity~\cite{Davi79}. A conservative estimate for this operator norm (e.g., for $\lino{A}_d^{[m]}$) is then given by:
			\begin{equation}
				\left\|\lino{A}\nolimits_d^{[m]}\right\|_{op}=\max_{\range{k}{1}{N_k}}\sqrt{\sum_{i=1}^{N_b^2} b^2_{ik}}\;;
			\end{equation}

			\item \textbf{Sum over} $N_k$ \textbf{channels}: the operator norm upper bound is $\sqrt{N_k}$, as a result of the triangular inequality applied over every pixel;
			\item \textbf{Shifting}: As shifting is assumed to be a one-to-one operation, the intensity values of each sample are unchanged, hence the operator norm is unitary;
			\item \textbf{Element-wise product by} $\mathbf{H}$: As pixels are scaled by the intensity value of $\mathbf{H}$, the operator norm is equal to the largest value of $\mathbf{H}$ (i.e., it is equal to one if $\mathbf{H}$ is a non-degenerate binary mask).
		\end{itemize}

\section{Proposed image reconstruction algorithm}
\label{sec:reconstruction}

    \subsection{Problem statement}
    \label{ssec:reconstruction_intro}

       	This section presents the proposed algorithm to produce an estimation $\widehat{\mathbf{X}}$ of the target image. Given the observation $\mathbf{y}$ of the optical device, we aim to make $\widehat{\mathbf{X}}$ as close as possible to the ideal (and typically unknown) reference $\mathbf{X}$ which generates it.
        We operate under the assumption that the observations are affected by noise modeled as an additive \gls{iid} Gaussian distribution with zero mean. While this hypothesis is common for previous Bayesian formulations of similar problems in the literature~\cite{Lonc15, Tan17a}, its validity is currently a point of contention in the scientific community, but is at least reasonable under sufficient high illumination~\cite{Cohn06}.

        The proposed \textbf{\gls{jodefu}} reconstruction algorithm is based on the following Bayesian formulation of the inverse problem:
        \begin{equation}
            \widehat{\mathbf{X}}=\arg\min_{\mathbf{X}} \frac{1}{2}\left\|\lino{A}(\mathbf{X})-\mathbf{y}\right\|^2_2+f(\mathbf{X})\,,
            \label{eq:jodefu_both}
        \end{equation}
        where the first term of the right side is the maximum a posteriori estimation, also known as data fidelity term, and $f(\mathbf{X}): \mathbb{R}^{N_i N_j \times N_k}\rightarrow \mathbb{R}^+$ is a regularization function, which we can customize according to our prior knowledge on the result to reconstruct and is used to counteract the ill-conditioned nature of the formulation~\cite{Hada02}. We focus our attention on the expression of the data fidelity term in Section~\ref{ssec:reconstruction_fidelity}, on the regularizer in Section~\ref{ssec:reconstruction_regularization}, and on the algorithm for solving this problem in Section~\ref{ssec:reconstruction_implementation}.

    \subsection{Data fidelity term}
    \label{ssec:reconstruction_fidelity}

        The \gls{jodefu} algorithm can be applied to the observation of any device that can be described within the \gls{mrca} image formation model. That is, other than the general \gls{mrca} itself, it may also approach the problem of sharpening (if $\lino{A}$ models a multiresolution sensing), that of demosaicing (if $\lino{A}$ models a mosaicing), and so on.
        While we specialize here on the \gls{mrca} framework, the proposed algorithm admits a solution for any image formation model, as long as we can define the properties of the operator $\lino{A}$ as a combination of those defined in Section~\ref{ssec:formation_properties}.

		Special care has to be taken in the sharpening scenario (and in general, for every setup in which $\lino{A}_+$ is an identity operator). As the observation is made of two separate acquisitions, the reconstruction problem is equivalent to:

		\begin{multline}
			\widehat{\mathbf{X}} = \arg\min_{\mathbf{X}}\frac{1}{2}\left\|\lino{A}\nolimits_d^{[p]}(\mathbf{X}) - \mathbf{P}\right\|_F\\
			+\frac{1}{2}\left\|\lino{A}\nolimits_d^{[m]}(\mathbf{X}) - \widetilde{\mathbf{M}}\right\|_F+ f(\mathbf{X})\,,
			\label{eq:reconstruction_pansharpening}
		\end{multline}
		which is very similar to the Bayesian formulation of the sharpening problem proposed in~\cite{Lonc15}, but assumes the same weight for the data fidelity term associated to the \gls{lri} and to the \gls{hri}.

        Previous works~\cite{Cham16} have shown that the algorithms that were proposed to solve image reconstruction problems based on multiresolution acquisition (e.g., pansharpening), are not well suited for the case of reconstruction of missing acquisitions (e.g., inpainting), which is also the case of the demosaicing as binary masks technically perform a subsampling.
        This motivates the need of an ad-hoc algorithm where these problems can be solved jointly.

        To deal with such scenarios, the proposed blur filter $\lino{A}_b$ is used to adjust the results even when the \gls{hri} is at the same scale of the reference. Its inclusion allows to recast our demosaicing problem from pure inpainting to an hybrid of inpainting/magnification. The formulation implies that some information from the suppressed pixels is contained in adjacent pixels, but this comes at the cost of reducing the spatial resolution of the final product.

	\subsection{Regularization}
	\label{ssec:reconstruction_regularization}

	    We want to setup here a \textit{proximal algorithm}. This class of algorithms works in very general conditions, allowing for some nonsmooth real-valued constraint on the cost function, and is relatively fast with respect to other alternatives~\cite{Pari14}. To this end, we propose a regularization function in the form:
	    \begin{equation}
	    	f(\mathbf{X})=\lambda\;g(\lino{L}(\mathbf{X}))\;,
	    \end{equation}
	    where we denote:
	    \begin{itemize}
	       	\item a \textbf{linear operator} $\lino{L}(\cdot): \mathbb{R}^{N_i N_j \times N_k}\rightarrow \mathtt{E}_l$, which describes $\mathbf{X}$ within a sparsity-inducing transformed normed space $\mathtt{E}_l$;
	       	\item a \textbf{metric function} $g(\cdot): \mathtt{E}_l \rightarrow \mathbb{R}^+$, for which it is possible to define a proximal operator $\prox_{\gamma g}(\cdot)$ scaled by a generic positive scalar $\gamma$;
	       	\item a \textbf{regularization parameter} $\lambda \in \mathbb{R}^+$, used to weight the contribute of the regularization with respect to the data term in the cost function; we sometimes refer to this term in its \textit{normalized form} $\overline{\lambda}=\lambda / \rho_y$, where $\rho_y$ is the dynamic range of the observation $\mathbf{y}$ (e.g., $\rho_y=255$ for 8 bits images).
	    \end{itemize}

	    For the linear transformation $\tens{W}=\lino{L}(\mathbf{X})$, the \gls{cassi} authors proposed to use a \textit{symlet-8 \glsentrylong{dwt}} and a \textit{\gls{dct}} transform in the spatial and spectral domain, respectively~\cite{Arce14}. In this work, we propose instead an approach based on the \textit{\glsreset{tv}\gls{tv}}, a regularizing transformation that acts as a discrete representation the \textit{\glsentrylong{rof} model}~\cite{Rudi92}. In its modern interpretation, the \gls{tv} is often seen as a sparsity-inducing operator working in the domain of image gradients. This favors piecewise constant images with sparse edges, which are typically a better representation of natural images~\cite{Cham16}.

	    Following the generic mathematical description of the third author~\cite{Cond17}, the \gls{tv}-based expression of $\tens{W}$ is a 4-way tensor $\tens{W}\in\mathbb{R}^{N_i \times N_j \times N_k \times N_m}$, whose forth dimension is made up of the gradients of $\tens{U}^{[x]}$. For the classic \gls{tv} in particular, where $N_m=2$, the gradients are taken along the horizontal and vertical spatial dimensions, and the elements $w_{ijkm}$ of $\tens{W}$ (assuming that the elements out of range in $\tens{U}^{[x]}$ are zero) are defined as follows:
	    \begin{equation}
	       	w_{ijkm}=
	       	\begin{cases}
	       		u^{[x]}_{ijk}-u^{[x]}_{i-1,j,k}&\textrm{for } m=1\\
	       		u^{[x]}_{ijk}-u^{[x]}_{i,j-1,k}&\textrm{for } m=2
	       	\end{cases}\,.
	       \label{eq:tv}
	    \end{equation}
	    Similar four way tensors can be also defined for alternative \gls{tv}-based operators, such as the \textit{\gls{utv}}~\cite{Cham11}, and the \textit{\gls{stv}}~\cite{Aber17}.

	    To define the metric function $g(\cdot)$, we took inspiration from the framework of the \textit{\gls{ctv}}~\cite{Dura16,Dura16a}, where $g(\cdot)$ is defined as a set of norms applied sequentially over different dimensions.

     	The most relevant that are also employed in this work, are defined below:
     	\begin{itemize}
     		\item $g(\tens{W})=\|\tens{W}\|_{p_1 p_2 p_3}$: this stands for the $\ell_{p_1}$, $\ell_{p_2}$, and $\ell_{p_3}$-norm applied, in this order, respectively to dimension of the gradient, that of the channels, and that of the pixels. Among those, the most widespread are the $\|\tens{W}\|_{221}$, which is used in the \textit{\glsentrylong{vtv}}~\cite{Bres08} and whose mathematical expression is:
     		\begin{equation}
     			\|\tens{W}\|_{221}=\sum_{i=1}^{N_i}\sum_{j=1}^{N_j}\sqrt{\sum_{k=1}^{N_k}\sum_{m=1}^{N_m}w^2_{ijkm}}
     			\label{eq:vtv}
     		\end{equation}
     		and the $\|\tens{W}\|_{111}$ norm, known as the \textit{\gls{lasso}}~\cite{Tibs11} and employed for the classic inversion of the \gls{cassi} acquisitions~\cite{Arce14}.

     		\item $g(\tens{W})=\|\tens{W}\|_{S_p \ell_q}$: this stands for the Shatten $p$-norm firstly applied on both the gradient and the bands' dimensions, and then the $\ell_q$-norm applied over the pixels; particularly good performances can be obtained with $\|\tens{W}\|_{S_1\ell_1}$, defined as:
     		\begin{equation}
     			\|\tens{W}\|_{S_1\ell_1}=\sum_{i=1}^{N_i}\sum_{j=1}^{N_j}\sqrt{\sum_{m=1}^{N_r}\xi_m^2\left(\mathbf{W}_{ij::}\right)}\,,
     			\label{eq:s1l1}
     		\end{equation}
     		where $\xi_m(\mathbf{W}_{ij::})$ is the $m$-th singular value and $N_r$ is the total amount of singular values of $\mathbf{W}_{ij::}$.
     	\end{itemize}

	\subsection{Implementation details}
	\label{ssec:reconstruction_implementation}

	  	The proposed \gls{jodefu} image reconstruction framework can be summarized as follows:
	  	\begin{equation}
	  		\widehat{\mathbf{X}}=\arg\min_{\mathbf{X}}\frac{1}{2}\|\lino{A}(\mathbf{X})-\mathbf{y}\|^2_2 + \lambda\;g(\lino{L}(\mathbf{X}))\,,
	  		\label{eq:jodefu}
	  	\end{equation}
		which is the minimization of a cost function composed by a differentiable data fidelity term and a regularization term whose metric function $g(\cdot)$ is a lower semi-continuous convex function.  As long as the adjoint operators for $\lino{A}$ and $\lino{L}$ are known, and it is possible to define a proximal operator for $g(\cdot)$, a variety of algorithms are available that iteratively converge to the desired solution $\widehat{\mathbf{X}}$. Those are known as \textit{proximal algorithms} in the literature~\cite{Pari14}. Among those, the \textit{Chambolle-Pock solver}~\cite{Cham10} is the most widespread, but we prefer here to employ instead the \textit{Loris-Verhoeven algorithm}~\cite{Lori11}, which simplifies the choice of the convergence parameters (as reported in~\cite{Cond22}).

		The full procedure, described by the Algorithm~\ref{alg:jodefu}, requires the definition of:
		\begin{itemize}
			\item the adjoint operator $\lino{A}^*$ and the operator norm $\|\lino{A}\|_{op}$ of $\lino{A}$, which were described in Section~\ref{ssec:formation_properties};
			\item the adjoint operator $\lino{L}^*$ and the operator norm $\|\lino{L}\|_{op}$ of $\lino{L}$: for orthogonal operators (such as the \gls{dct} and some wavelets), their operator norm is unitary. For the classic \gls{tv} the generic $v_{ijk}$ element of $\tens{V}\in\mathbb{R}^{N_i \times N_j \times N_k}$, of $\lexi(\tens{V})=\lino{L}^*(\tens{W})$ is:
			\begin{equation}
				v_{ijk} = (w_{i,j,k,1} - w_{i-1,j,k,1}) + (w_{i,j,k,1} - w_{i,j-1,k,2})\,,
			\end{equation}
			where we assume once again that the elements out of range of $\tens{W}$ are null. Its operator norm is $\|\lino{L}\|_{op}=\sqrt{8}$~\cite{Cond17}. For the other \gls{tv}-like operators, we redirect the reader towards the related articles~\cite{Cham11, Aber17}.
			\item the scaled \textit{proximal operator} $\prox_{\lambda g\fenc}(\cdot)$ of the Fenchel conjugate of $g(\cdot)$: a summarizing table of its expression is provided in~\cite{Dura16a}. We just remind here that, for the $\|\tens{W}\|_{221}$ norm, this is equal to:
			\begin{equation}
				\prox\nolimits_{\lambda g\fenc}(\mathbf{y})= \frac{\tens{W}}{\max\left(\sqrt{\frac{1}{\lambda}\sum\limits_{k=1}^{N_k}\sum\limits_{m=1}^{N_m} \mathbf{W}_{::km}} \;,\; 1\right)}\;,
				\label{eq:proximal}
			\end{equation}
			where $\max(\mathbf{X},\; 1)$ is an operator substituting with $1$ all elements of $\mathbf{X}$ that are superior to 1, and the fraction stands for an element-wise division broadcasted over 4 dimensions.

		\end{itemize}

		\begin{algorithm}
    \DontPrintSemicolon
    \KwResult{ Estimated product $\widehat{\mathbf{X}}$ }
    
    \SetKwData{Left}{left}\SetKwData{This}{this}\SetKwData{Up}{up}%
    \SetKwInOut{KwInput}{Input}
    \SetKwInOut{KwPreproc}{Pre-processing}
    \SetKwInOut{KwInitialization}{Initialization}
    
    \textbf{Input}:
        \begin{itemize}
            \item Acquisition: $\mathbf{y}\in\mathbb{R}^{N_iN_j}$ (with the \gls{lri} samples histogram matched to the \gls{hri})
            \item Direct model operator $\lino{A}(\cdot)$, according to eq.~\eqref{eq:mrca}, with $\lino{A}^*$ and $\|\lino{A}\|_{op}$ defined in Section~\ref{ssec:formation_properties};
            \item Domain transformation operator $\lino{L}(\cdot)$, e.g. a \gls{tv} according to eq.~\eqref{eq:tv}, an \gls{utv}, or a \gls{stv}, with $\lino{L}^*$ and $\|\lino{L}\|_{op}$ defined in Section~\ref{ssec:reconstruction_implementation};
        	\item Proximal operator $\prox_{\lambda g\fenc}(\cdot)$, e.g. from eq.~\eqref{eq:proximal};
            \item Regularization parameter: $\lambda$ (default: $10^{-3} \rho_y$, where  $\rho_y$ is the dynamic range of $\mathbf{y}$);
            \item Over-relaxation parameter: $\rho_o$ (default: $1.9$);
            \item Maximum number of iterations: $q^{[max]}$ (default: $250$);
        \end{itemize}
    
    \textbf{Initialization}:
        \begin{itemize}
            \item First convergence parameter: $\tau=0.99/\|\lino{A}\|^2_{op}$
            \item Second convergence parameter: $\sigma=1/(\tau\|\lino{L}\|^2_{op})$
            \item Primal variable: $\mathbf{X}^{(0)}=\lino{A}^*(\mathbf{y})$
            \item Dual variable: $\tens{W}^{(0)}=\lino{L}\left(\mathbf{X}^{(0)}\right)$
            \item Iteration: $q=0$
        \end{itemize}
    
    \While{$q < q^{[max]}$}{
    	$\mathbf{V}^{(q)}=\lino{A}^*\left(\lino{A}\left(\mathbf{X}^{(q)}\right)-\mathbf{y}\right)$\;
    	$\mathbf{X}^{\left(q+\frac{1}{2}\right)}=\mathbf{X}^{(q)}-\tau\left(\mathbf{V}^{(q)}+ \lino{L}^*\left(\tens{W}^{(q)}\right)\right)$\;
        $\tens{W}^{\left(q+\frac{1}{2}\right)} = \prox_{\lambda g\fenc}\left(\tens{W}^{(q)}+\sigma \lino{L}\left(\mathbf{X}^{\left(q+\frac{1}{2}\right)}\right)\right)$\;
        $\mathbf{X}^{(q+1)}= \mathbf{X}^{(q)}-\rho_o\tau\left(\mathbf{V}^{(q)}+\lino{L}^*\left(\tens{W}^{\left(q+\frac{1}{2}\right)}\right)\right)$\;
        $\tens{W}^{(q+1)}=\tens{W}^{(q)}+\rho_o\left(\tens{W}^{(q+\frac{1}{2})}-\tens{W}^{(q)}\right)$\;
        $q \gets q+1$\;
     }
    \Return $\widehat{\mathbf{X}}=\mathbf{X}^{(q^{[max]})}$\;

\caption{\Gls{jodefu} algorithm, based on the Loris-Verhoeven solver~\cite{Lori11} with over-relaxation~\cite{Cond22}.}
\label{alg:jodefu}
\end{algorithm}

		The \gls{jodefu} framework allows for multiple ways to construct a custom cost function, which include the choice for the regularization parameter $\lambda$, for the metric function $g(\cdot)$, and the linear transformation operator $\lino{L}$. We propose here two possible solutions, whose specifics are shown in table~\ref{tab:jodefu_variants}: the \textbf{\gls{jodefu} v1} provides reasonable performances while keeping the computation time relatively short, while the \textbf{\gls{jodefu} v2} variant produces optimal performances if there is no constraint on time.

		Since the proposed algorithm is iterative in nature, the computational burden is surely higher than the classic one-shot algorithms, which may impact the scalability of the proposed algorithm. In terms of computational complexity, the linear operators $\lino{A}$ and $\lino{L}$ used in the proposed algorithm scale linearly with the number of samples $N_s=N_iN_jN_k$, being composed of convolutions with a spatial kernel or of weighted element-wise products.
		The proximal operator is the main bottleneck for the v2 version of the algorithm, since the computational complexity of eq.~\eqref{eq:s1l1} is $\mathcal{O}(N_k N_m \min(N_m,\,N_k))$ for each pixel, as it involves a \glsentrylong{svd}~\cite{Golu13}, while the v1 is limited to a $\mathcal{O}(N_k N_m)$ complexity from eq.~\eqref{eq:vtv}.

		\begin{table}
	\caption[\glsentryshort{jodefu} framework variants]{Suggested setups for \gls{mrca} image reconstruction algorithms.
	}
	
	\begin{adjustbox}{width=\linewidth}
		\begin{tabular}{l|c|c|c}
			
			\hline
			\multicolumn{4}{c}{\textbf{\glsentryshort{jodefu} setups}}\\
			\hline
			&$g(\tens{W})$&$\lino{L}(\cdot)$&$\lino{A}_b(\cdot)$\\
			\hline
			\textbf{\glsentryshort{jodefu} v1}&$\|\tens{W}\|_{221}$~\cite{Dura16}&\glsentryshort{tv}~\cite{Rudi92}&Identity\\
			\textbf{\glsentryshort{jodefu} v2}&$\|\tens{W}\|_{S_2\ell_1}$~\cite{Dura16}&\glsentryshort{utv}~\cite{Cham11} or \glsentryshort{stv}~\cite{Aber17}&$\rho_b=1\text{--}1.5$ \si{px}\\
			\hline
			\multicolumn{4}{c}{\textbf{Cascaded classic algorithms setups}}\\
			\hline
			&\textbf{\Glsentryshort{pan}
				interpolation}&\textbf{Demosaicing}&\textbf{Sharpening}\\
			\hline
			\textbf{Classic v1}&\glsentryshort{tps-rbf}~\cite{Buhm03}&\glsentryshort{ari}~\cite{Monn17}&\glsentryshort{mtf-glp-hpm}~\cite{Aiaz06}\\
			\textbf{Classic v2}&\glsentryshort{tps-rbf}~\cite{Buhm03}&\glsentryshort{id}~\cite{Miho17}&\glsentryshort{gsa}~\cite{Aiaz07}\\
			\hline
		\end{tabular}
	\end{adjustbox}\\

	\scriptsize{The first half refers to the \gls{jodefu} algorithm, while the rows labeled with classic v1 and v2 refer to the alternative algorithms described in Section~\ref{ssec:experiments_reconstruction}.}
	\label{tab:jodefu_variants}
\end{table}

\section{Experiments}
\label{sec:experiments}

	In this section, the proposed image formation/reconstruction framework is tested under different viewpoints:
	\begin{itemize}
		\item In Section~\ref{ssec:experiments_formation}, we analyze the acquisition obtained with a variety of image formation methods. For each acquisition under test, we compare the reconstructed products obtained both with a representative state-of-the-art classic reconstruction algorithm and with the proposed \gls{jodefu} algorithm. This experiment aims to show the effectiveness of the \gls{mrca} in modeling a wide variety of capturing techniques, by verifying that the quality of the reconstructed image meets the standards set up by the previous literature when this model is employed within the proposed reconstruction algorithm;
		\item in Section~\ref{ssec:experiments_reconstruction}, the image formation is fixed to the proposed \gls{mrca} and we inspect a variety of solutions for the image reconstruction, by comparing the results when a different amount of \gls{lri} channels are encoded in the acquisition. This experiments aims to show the capability of the \gls{jodefu} in recovering the relevant information embedded in the compressed acquisition under different conditions.
		\item in Section~\ref{ssec:experiments_parameters}, we analyze the effect of the parameters associated to the \gls{jodefu} algorithm. The experiment aims to test the robustness of the proposed reconstruction algorithm with respect to deviations from an ideal parametric setup.
	\end{itemize}
	To introduce the experiments, a description of the employed datasets, of the experimental setup, and of the validation method is given in Section~\ref{ssec:experiments_setup}.
	The experiments provided in this section are fully reproducible with the MATLAB implementation of the algorithm\footnote{[Online]. Available: \href{https://github.com/danaroth83/jodefu}{https://github.com/danaroth83/jodefu}} and additional results are available in the \textit{supplementary materials}.

    \subsection{Experimental setup}
    \label{ssec:experiments_setup}

    	Our validation setup consists of four steps:
    	\begin{itemize}
    		\item \textbf{Reference choice}: where we select a high resolution image as reference $\mathbf{X}\in\mathbb{R}^{N_i N_j \times N_k}$, that is referred as \gls{gt};
    		\item \textbf{Simulation}: where the acquisition $\mathbf{y}=\lino{A}(\mathbf{X})$ is evaluated from the direct model $\lino{A}$ under test (e.g., with the architecture in \figurename~\ref{fig:direct} for the \gls{mrca});
    		\item \textbf{Testing}: where the reconstruction algorithm under test is applied to the observation, in order to achieve an estimation $\widehat{\mathbf{X}}$ of $\mathbf{X}$;
    		\item \textbf{Comparison}: where the estimated product $\widehat{\mathbf{X}}$ and the reference $\mathbf{X}$ are compared by evaluating a series of quality indices.
    	\end{itemize}
    
        \begin{table}
    \centering
    \caption[Characteristics of remotely sensed datasets]{Characteristics of the \gls{gt} of the datasets employed in the tests of Section~\ref{sec:experiments}.}
    \begin{adjustbox}{width=\linewidth}
	    \begin{tblr}{
	    	colspec={l|c|c|c|c},
    		row{5-6}={fg=added},
    	}	
    		\hline
	        \textbf{Label}&\textbf{Scene}&\textbf{Sensor}&\textbf{\glsentryshort{gsd}}&\textbf{Sizes} [\si{px}]\\
	        \hline
	        \textbf{Beijing}&Bird's nest, China&\glsentryshort{wv2}&1.6 \si{m}&$512 \times 512$\\
	        \textbf{Janeiro}&Bay area, Brazil&\glsentryshort{wv3}&1.2 \si{m}&$512 \times 512$\\
	        \textbf{Washington}&Capitol building, U.S.&\glsentryshort{wv3}&1.6 \si{m}&$512 \times 512$\\
	        \textbf{Stockholm}&Central city, Sweden&\glsentryshort{wv3}&1.2 \si{m}&$512 \times 512$\\
	        \textbf{Fields}&Mineral mountain, China&\glsentrylong{iko}&3.2 \si{m}&$256 \times 256$\\
	        \hline
	    \end{tblr}
    \end{adjustbox}\\\vspace{0.1 cm}\hspace{-0.6cm}
	\scriptsize{The \gls{gsd} refers to the spatial resolution of the \gls{gt}.}
    \label{tab:datasets_reduced}
\end{table}

    	Each reference dataset is composed of a \gls{hri}/\gls{lri} image bundle acquired almost simultaneously, originally featuring a scale ratio of $1:4$, although the tests are performed at reduced resolution with a scale ratio of $\ratio=2$. The \gls{hri} is monochromatic (i.e. a \gls{pan}) and the \gls{lri} has up to 8 channels.

    	The bundles were acquired by the \glsentrylong{iko}, \gls{wv2} and \gls{wv3} satellites, and are available for download on the MAXAR Technologies website~\cite{web_Maxa}. Their characteristics are shown in \tablename~\ref{tab:datasets_reduced}. Additional experiments over different datasets are available in the supplementary materials.

    	For the simulation step, when applicable, the \gls{hri} is given as a spatial degradation of a \gls{hri} at higher resolution, instead of spectral degradation of the \gls{gt}, in order to follow the Wald's protocol for reduced resolution validation~\cite{Wald97}.

    	In the spectral degradation model $\lino{A}_d^{[p]}$, the weighting coefficients are always set as equal to $w_{1,k}=1/N_k$, for all $\range{k}{1}{N_k}$, so that the \gls{hri} is modeled as the average of the channels of the \gls{gt}. In the spatial degradation model $\lino{A}_c^{[p]}$, the blurring kernels $\mathbf{B}$ are Gaussian functions whose cutoff frequency matches the one of the \gls{mtf} of the sensors. The $\lino{A}_b(\cdot)$ is implemented by a \gls{iir} Butterworth filter whose bilateral cutoff frequency is $1/\rho_b$, with $\rho_b$ denoting the diameter of the blurring filter and expressed in pixels (\si{px}). All \gls{jodefu} algorithms are run for $250$ iterations. In the \gls{jodefu} preprocessing stage, the overall mean and \glsentrylong{std} of the \gls{lri} samples are equalized to that of \gls{hri} samples.

        For the comparison step, we employ as quality indices the \textit{\gls{psnr}}, the \textit{\gls{sam}}~\cite{Yuha92}, and the \textit{\gls{ssim}}~\cite{Wang04}, which is given as average over all bands.

        \begin{figure}
    \captionsetup[subfigure]{justification=centering}
    \centering
    
    \begin{subfigure}[t]{0.32\linewidth}
        \includegraphics[trim={10 10 10 10}, clip, width=0.99\linewidth]{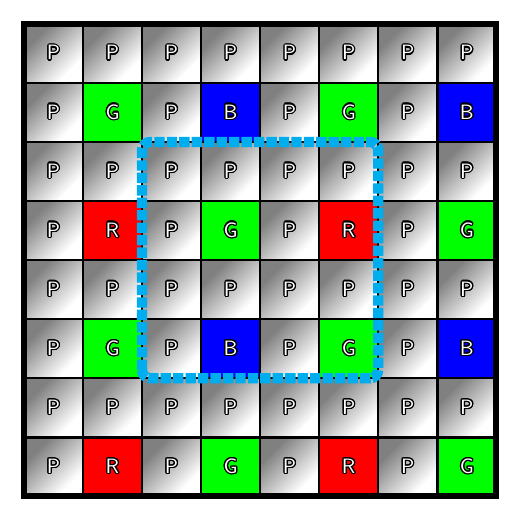}
        \caption{Bayer (3 Bands)~\cite{Baye76}}
        \label{fig:cfa_pattern_bayer}
    \end{subfigure}
    \hfil
    \begin{subfigure}[t]{0.32\linewidth}
        \includegraphics[trim={10 10 10 10}, clip, width=0.99\linewidth]{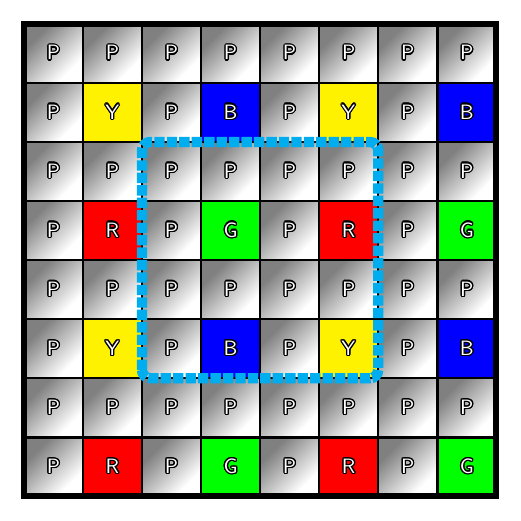}
        \caption{Periodic (4 Bands)~\cite{Miao06}}
        \label{fig:cfa_pattern_ubt4}
    \end{subfigure}
    \hfil
    \begin{subfigure}[t]{0.32\linewidth}
        \includegraphics[trim={10 10 10 10}, clip, width=0.99\linewidth]{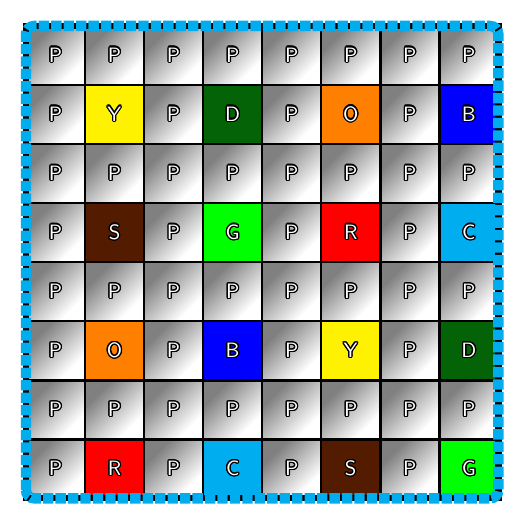}
        \caption{Uniform Binary Tree (8 bands)~\cite{Miao06}}
        \label{fig:cfa_pattern_ubt8}
    \end{subfigure}
    
    \caption[Combined \glsentryshort{pan} and \glsentryshort{ms} \glsentryshort{cfa} mask patterns]{Masks used for the experiments of Section~\ref{ssec:experiments_reconstruction}. The masks are obtained by mosaicing a set of \gls{hri} sensors (in gray) to classical literature designs for \gls{lri} masks, color coded with their characteristic channel. The dashed outline denotes the periodicity of each mask.}
    \label{fig:cfa_pattern}
\end{figure}

    \subsection{Image formation}
    \label{ssec:experiments_formation}

\begin{table*}
	\caption{Results of classic and proposed reconstruction algorithms for different methods of image formation with the 4-bands\\``Washington'' and ``Fields'' dataset.
	}
	\begin{adjustbox}{width=\linewidth}
		\begin{tblr}{
			colspec={l|c|l|c|ccc|c|ccc},
		}
			\hline
			\multirow{2}{*}{\textbf{Image formation}}&\multirow{2}{*}{$\rho_c$}&\multirow{2}{*}{\textbf{Image reconstruction}}&\multicolumn{4}{|c|}{Washington}&\multicolumn{4}{|c}{Fields}\\
			\cline{4-11}
			&&&$\overline{\lambda}$&\textbf{\glsentryshort{ssim}}&\textbf{\glsentryshort{psnr}}&\textbf{\glsentryshort{sam}}&$\overline{\lambda}$&\textbf{\glsentryshort{ssim}}&\textbf{\glsentryshort{psnr}}&\textbf{\glsentryshort{sam}}\\
			\hline
			Reference (\glsentryshort{gt})&1.000&-&-&1&$\infty$&0&-&1&$\infty$&0\\
			\hline
			\multirow{2}{*}{Multiresolution sensing}&\multirow{2}{*}{0.500}&Pansharpening~\cite{Aiaz12}&-&0.9772&31.14&3.751&-&0.9724&41.42&2.483\\
			&&\glsentryshort{jodefu} v1&$1\times10^{-3}$&\textbf{0.9868}&\textbf{33.22}&\textbf{2.797}&$2\times10^{-4}$&\textbf{0.9744}&\textbf{42.10}&\textbf{2.313}\\
			\hline
			\multirow{2}{*}{Mosaicing}&\multirow{2}{*}{0.250}&Demosaicing~\cite{Miho15}&-&\textbf{0.9613}&\textbf{30.38}&\textbf{4.361}&-&\textbf{0.9513}&\textbf{38.92}&\textbf{3.465}\\
			&&\glsentryshort{jodefu} v1&$2\times10^{-3}$&0.9312&28.36&4.371&$8\times10^{-4}$&0.9450&38.54&3.276\\
			\hline
			\multirow{2}{*}{\glsentryshort{cassi} acquisition~\cite{Arce14}}&\multirow{2}{*}{0.251}&\glsentryshort{cassi} reconstruction~\cite{Arce14}&$3\times 10^{-3}$&0.7645&24.48&9.750&$2\times 10^{-3}$&0.8717&33.75&7.795\\
			&&\glsentryshort{jodefu} v1&$2\times 10^{-3}$&\textbf{0.8600}&\textbf{26.89}&\textbf{5.954}&$1\times 10^{-3}$&\textbf{0.9147}&\textbf{36.66}&\textbf{4.569}\\
			\hline
			\multirow{3}{*}{\glsentryshort{mrca}}&\multirow{3}{*}{0.250}&Classic v2 (\tablename~\ref{tab:jodefu_variants})&-&0.9156&28.13&6.246&-&0.9517&\textbf{39.54}&3.503\\
			&&\glsentryshort{jodefu} v1&$1\times 10^{-3}$&0.9446&29.34&4.553&$1\times 10^{-3}$&0.9442&38.09&3.652\\
			&&\glsentryshort{jodefu} v2 with UTV ($\rho_b=1.4$)&$1\times 10^{-3}$&\textbf{0.9560}&\textbf{29.79}&\textbf{3.841}&$4\times 10^{-4}$&\textbf{0.9541}&39.19&\textbf{3.035}\\
			\hline
		\end{tblr}
	\end{adjustbox}\\

	\hspace{0.2cm}Best results for each image formation method are shown in bold.
	\label{tab:compression}	
\end{table*}
		\begin{figure*}
	\captionsetup[subfigure]{justification=centering}
	\centering
	\begin{subfigure}[t]{0.21\linewidth}
		\includegraphics[width=\linewidth]{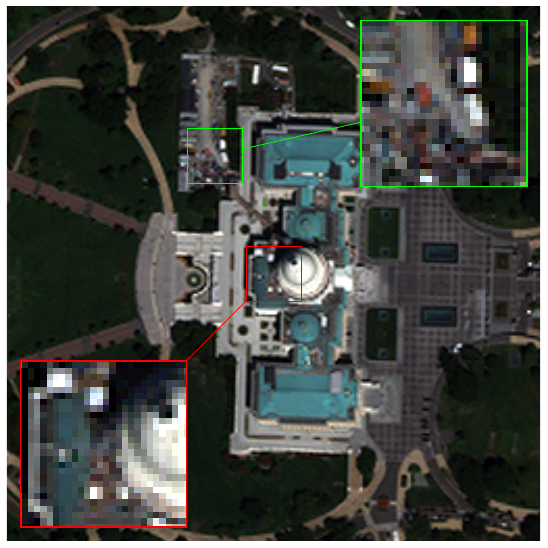}
		\caption{Reference (\glsentryshort{gt})}
		\label{fig:washington_4_gt}
	\end{subfigure}
		\hfil
	\begin{subfigure}[t]{0.21\linewidth}
		\includegraphics[width=\linewidth]{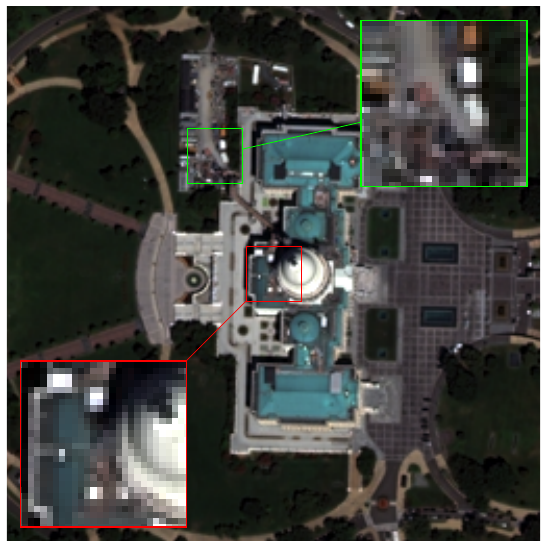}
		\caption{Multires. sens. - Class.~\cite{Aiaz06}}
		\label{fig:washington_4_pansharpening_classic}
	\end{subfigure}
	\hfil
	\begin{subfigure}[t]{0.21\linewidth}
		\includegraphics[width=\linewidth]{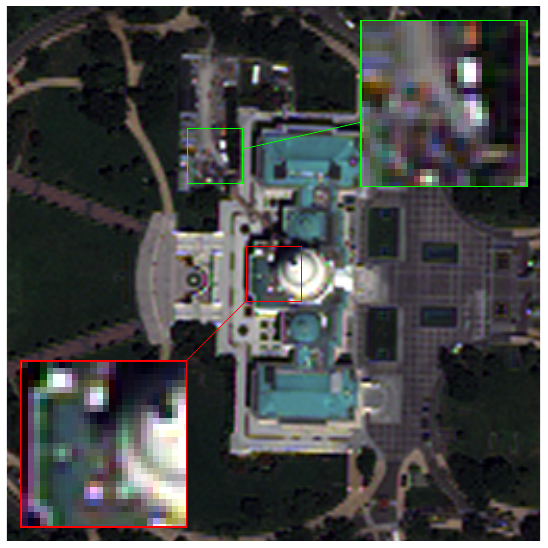}
		\caption{Mosaicing - Classic~\cite{Miho17}}
		\label{fig:washington_4_demosaicing_classic}
	\end{subfigure}
	\hfil
	\begin{subfigure}[t]{0.21\linewidth}
		\includegraphics[width=\linewidth]{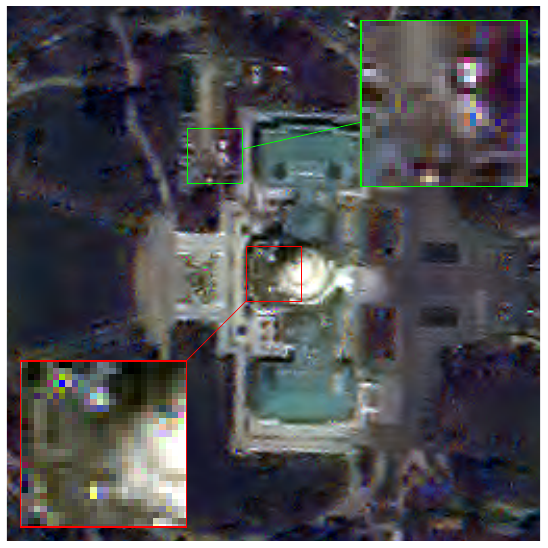}
		\caption{\glsentryshort{cassi} - Classic~\cite{Arce14}}
		\label{fig:washington_cassi_classic}
	\end{subfigure}

	\smallskip
	
	\begin{subfigure}[t]{0.21\linewidth}
		\includegraphics[width=\linewidth]{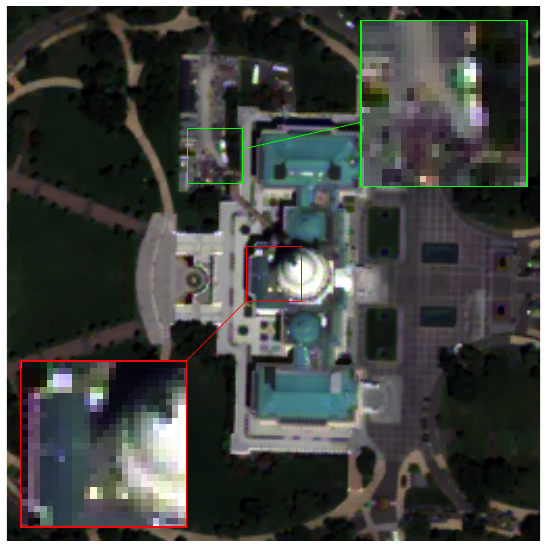}
		\caption{\glsentryshort{mrca} - \glsentryshort{jodefu} v1}
		\label{fig:washington_4_mrca}
	\end{subfigure}
	\hfil
	\begin{subfigure}[t]{0.21\linewidth}
		\includegraphics[width=\linewidth]{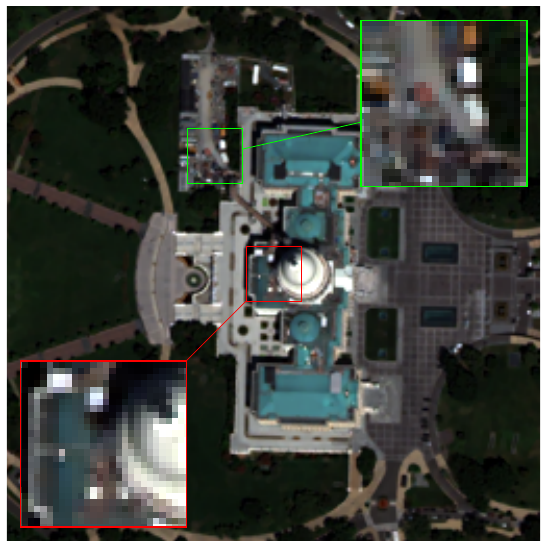}
		\caption{Multires. sens. - \glsentryshort{jodefu} v1}
		\label{fig:washington_4_pansharpening_jodefu}
	\end{subfigure}
	\hfil
	\begin{subfigure}[t]{0.21\linewidth}
		\includegraphics[width=\linewidth]{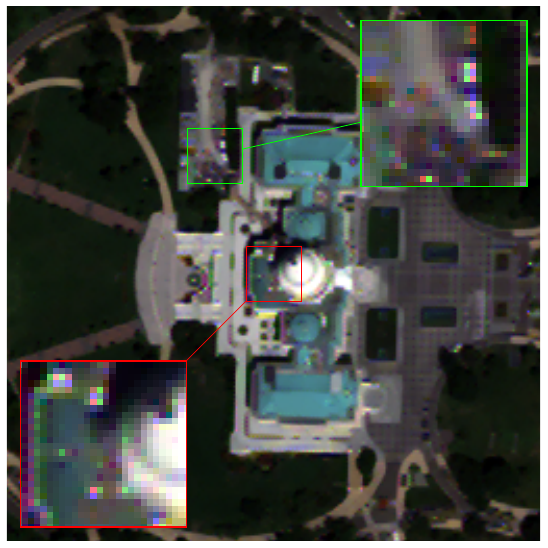}
		\caption{Mosaicing - \glsentryshort{jodefu} v1}
		\label{fig:washington_4_demosaic_jodefu}
	\end{subfigure}
	\hfil
	\begin{subfigure}[t]{0.21\linewidth}
		\includegraphics[width=\linewidth]{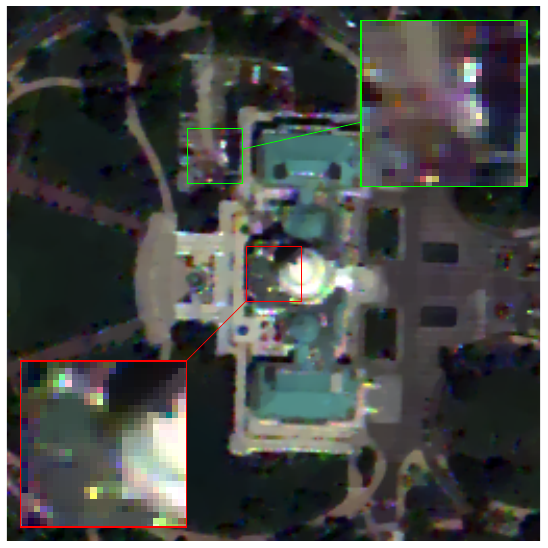}
		\caption{\glsentryshort{cassi} - \glsentryshort{jodefu} v1}
		\label{fig:washington_4_cassi_jodefu}
\end{subfigure}

	\caption[Image formation comparison]{Visual comparison for different image formation methods, comparing the reconstructed product obtained with the proposed \gls{jodefu} algorithm and with classic solutions available in the literature. The images show a $256 \times 256$ cropped area of the 4-band ``Washington'' dataset. Zoomed areas are provided for a detailed comparison.}
	\label{fig:washington_compression}
\end{figure*}

    	We consider here a selection of image formation methods and compare the obtained quality of the estimated product, produced both with a carefully selected classic image reconstruction algorithm and with the \gls{jodefu}. For completeness, this approach is also applied to an observation obtained with the complete declination of the \gls{mrca} method, using the mask of \figurename~\ref{fig:cfa_pattern_ubt4}.

    	Since formation methods can be interpreted as compressed acquisitions, the final observation generally contains less samples than the datacube to reconstruct. Therefore, their \textit{compression ratio} $\rho_c$ can be defined as the ratio between the amount of samples of the observation $\mathbf{y}$ and the reconstructed product $\widehat{\mathbf{X}}$. For example, if the dynamic range of all samples is the same, the compression ratio of the full \gls{mrca} model is equal to:
    	\begin{equation}
    		\rho_c=\frac{N_i N_j}{N_i N_j N_k}=\frac{1}{N_k}\,,
    	\end{equation}
    	as the multiresolution setup compresses the acquisition over a single \gls{fpa} matching the spatial dimensions of the reconstructed product.

    	We investigate the follow image formation methods:
    	\begin{itemize}
    		\item \textbf{Multiresolution sensing}: with the same scale ratio of the \gls{mrca}, this is compared with the \textit{\gls{mtf-glp-hpm}} algorithm~\cite{Aiaz12}, the best performing one for this dataset among the classic ones that were tested;
    		\item \textbf{Mosaicing}: applying a 4-band mask with period $2 \times 2$, which also reaches the same compression ratio of the \gls{mrca}, which is compared with the \textit{\gls{id}} demosaicing algorithm;
    		\item \textbf{\gls{cassi}}: using the model proposed by the authors employing one acquisition in their single dispersion variant and comparing with the proposed algorithm for the reconstruction~\cite{Arce14}.
    	\end{itemize}

    	The validation procedure is applied to the ``Washington'' and ``Fields'' datasets and compared to the baseline \gls{jodefu}, with the results shown in \tablename~\ref{tab:compression} and a visual comparisons in \figurename~\ref{fig:washington_compression}.

\begin{table*}
	\renewcommand\arraystretch{1.4}
	\centering
	\caption[\glsentryshort{mrca} reconstruction]{Results of \gls{mrca} image reconstruction with different amount of mosaiced bands for the ``Janeiro'' and ``Stockholm'' dataset.
	}

	\begin{adjustbox}{width=0.85\linewidth}	
		\begin{tblr}{
			colspec={c|l|ccc|ccc|ccc}
		}
			\hline
			&&\multicolumn{3}{c|}{\textbf{3 bands (\glsentryshort{rgb})}}&\multicolumn{3}{c|}{\textbf{4 bands (\glsentryshort{rgb} + \glsentryshort{nir})}}&\multicolumn{3}{c}{\textbf{8 bands (All \glsentryshort{vis}/\glsentryshort{nir})}}\\
			\hline
			&&\glsentryshort{ssim}&\glsentryshort{psnr}&\glsentryshort{sam}&\glsentryshort{ssim}&\glsentryshort{psnr}&\glsentryshort{sam}&\glsentryshort{ssim}&\glsentryshort{psnr}&\glsentryshort{sam}\\
			\hline
			&\textbf{Reference (\glsentryshort{gt})}&1&$\infty$&0&1&$\infty$&0&1&$\infty$&0\\
			\hline
			\multirow{5}{*}{\rotatebox{90}{Janeiro}}&\textbf{Classic v1}&\textbf{0.9634}&\textbf{32.30}&\textbf{2.502}& - & - & - & - & - & - \\
			&\textbf{Classic v2}&0.8680&28.77&5.253&0.9049&28.11&6.463&0.8917&28.50&9.166\\
			\cline{2-11}
			&\textbf{\glsentryshort{jodefu} v1} ($\overline{\lambda}=2\times10^{-3}$)&0.8803&28.78&4.988&0.9159&28.38&5.218&0.9080&29.08&7.220\\
			&\textbf{\glsentryshort{jodefu} v2 with \glsentryshort{utv}}&0.9050&29.57&3.901&\textbf{0.9264}&\textbf{28.58}&\textbf{4.891}&\textbf{0.9247}&\textbf{29.63}&\textbf{6.693}\\
			\cline[dashed]{2-11}
			&\textbf{\glsentryshort{jodefu} v2 options}&\multicolumn{3}{c|}{$\overline{\lambda}=2\times10^{-3},\;\rho_b=1$}&\multicolumn{3}{c|}{$\overline{\lambda}=1\times10^{-3},\;\rho_b=1.4$}&\multicolumn{3}{c}{$\overline{\lambda}=1\times10^{-3},\;\rho_b=1.4$}\\
			\hline
			\multirow{5}{*}{\rotatebox{90}{Stockholm}}&\textbf{Classic v1}&\textbf{0.9765}&\textbf{32.33}&\textbf{2.552}& - & - & - & - & - & - \\
			&\textbf{Classic v2}&0.9064&29.41&4.530&0.9240&28.77&7.512&0.8838&27.32&11.08\\
			\cline{2-11}
			&\textbf{\glsentryshort{jodefu} v1} ($\overline{\lambda}=2\times10^{-3}$)&0.9050&29.05&4.196&0.9249&28.54&6.075&0.8905&27.47&8.749\\
			&\textbf{\glsentryshort{jodefu} v2 with \glsentryshort{utv}}&0.9241&29.75&3.255&\textbf{0.9414}&\textbf{29.30}&\textbf{5.536}&\textbf{0.9105}&\textbf{28.17}&\textbf{8.181}\\
			\cline[dashed]{2-11}
			&\textbf{\glsentryshort{jodefu} v2 options}&\multicolumn{3}{c|}{$\overline{\lambda}=2\times10^{-3},\;\rho_b=1.4$}&\multicolumn{3}{c|}{$\overline{\lambda}=2\times10^{-3},\;\rho_b=1.4$}&\multicolumn{3}{c}{$\overline{\lambda}=1\times10^{-3},\;\rho_b=1.4$}\\
			\hline
		\end{tblr}
	\end{adjustbox}\\

	\vspace{0.2cm}

	\hspace{0.5cm}\footnotesize{The image construction is obtained with the \gls{mrca} model using the masks shown in \figurename~\ref{fig:cfa_pattern}. Best results are marked in bold fonts.}
	\label{tab:janeiro_bands}
\end{table*}
		\begin{figure*}
	\captionsetup[subfigure]{justification=centering}
	\centering
	\begin{subfigure}[t]{0.21\linewidth}
		\includegraphics[width=\linewidth]{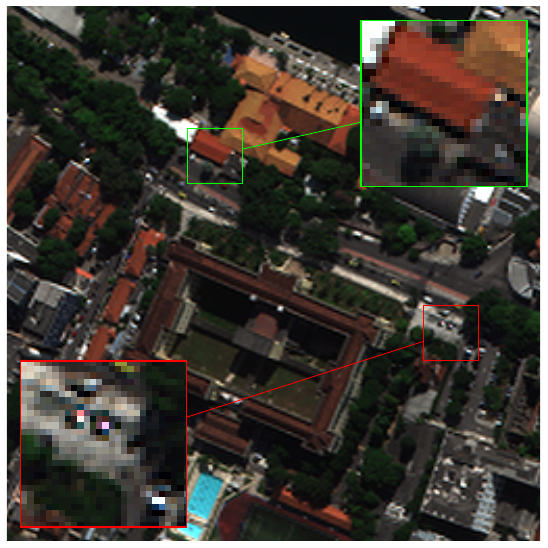}
		\caption{Reference (\glsentryshort{gt})}
		\label{fig:janeiro_gt}
	\end{subfigure}
	\hfil
	\begin{subfigure}[t]{0.21\linewidth}
		\includegraphics[width=\linewidth]{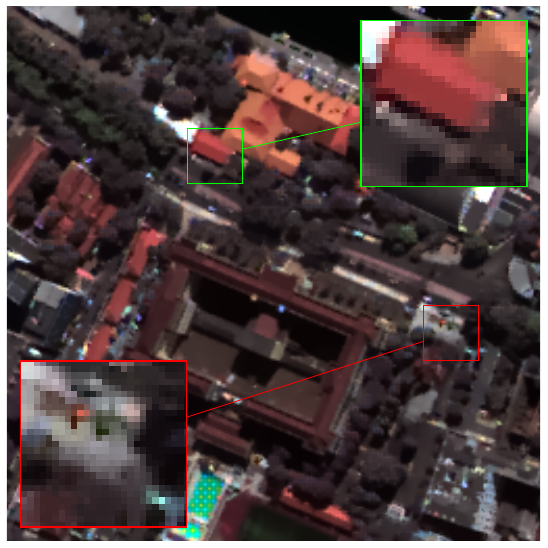}
		\caption{3 bands - \glsentryshort{jodefu} v2}
		\label{fig:janeiro_3_jodefu}
	\end{subfigure}
	\hfil
	\begin{subfigure}[t]{0.21\linewidth}
		\includegraphics[width=\linewidth]{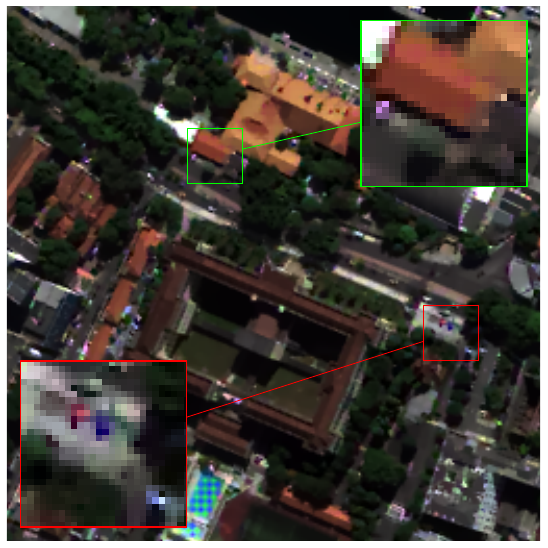}
		\caption{4 bands - \glsentryshort{jodefu} v2}
		\label{fig:janeiro_4_jodefu}
	\end{subfigure}
	\hfil
	\begin{subfigure}[t]{0.21\linewidth}
		\includegraphics[width=\linewidth]{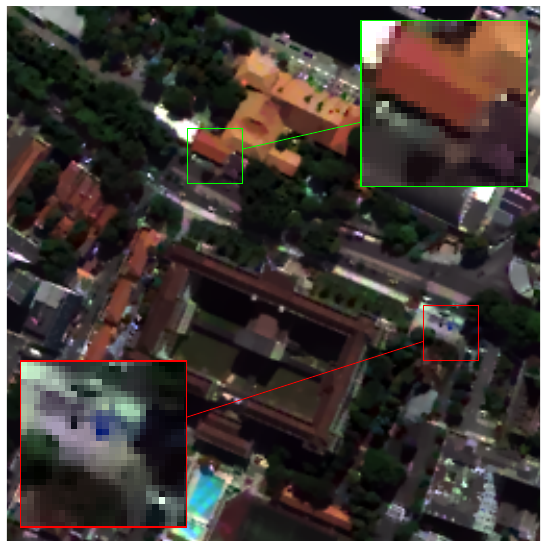}
		\caption{8 bands - \glsentryshort{jodefu} v2}
		\label{fig:janeiro_8_jodefu}
	\end{subfigure}
	
	\smallskip
	
	\begin{subfigure}[t]{0.21\linewidth}
		\includegraphics[width=\linewidth]{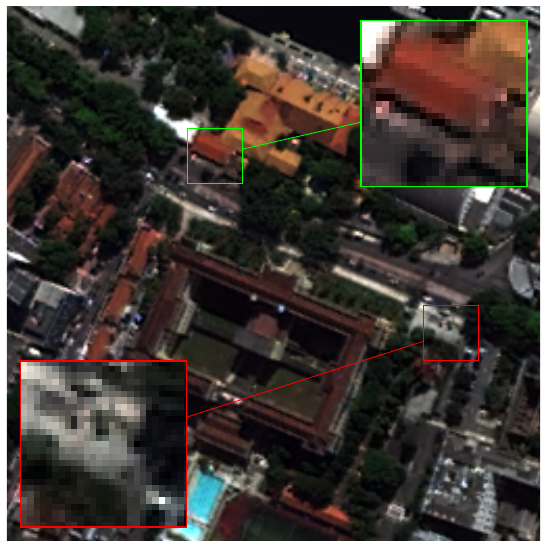}
		\caption{3 bands - Classic v1}
		\label{fig:janeiro_3_classic_v1}
	\end{subfigure}
	\hfil
	\begin{subfigure}[t]{0.21\linewidth}
		\includegraphics[width=\linewidth]{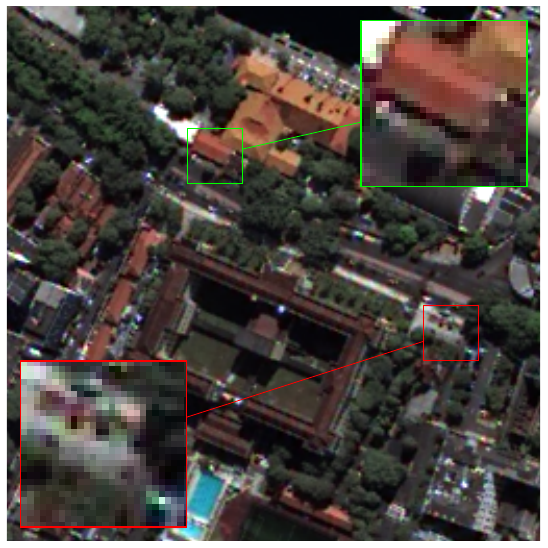}
		\caption{3 bands - Classic v2}
		\label{fig:janeiro_3_classic_v2}
	\end{subfigure}
	\hfil
	\begin{subfigure}[t]{0.21\linewidth}
		\includegraphics[width=\linewidth]{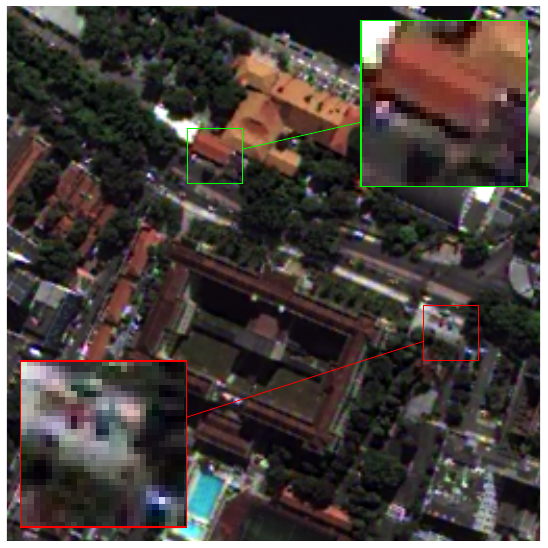}
		\caption{4 bands - Classic v2}
		\label{fig:janeiro_4_classic_v2}
	\end{subfigure}
	\hfil
	\begin{subfigure}[t]{0.21\linewidth}
		\includegraphics[width=\linewidth]{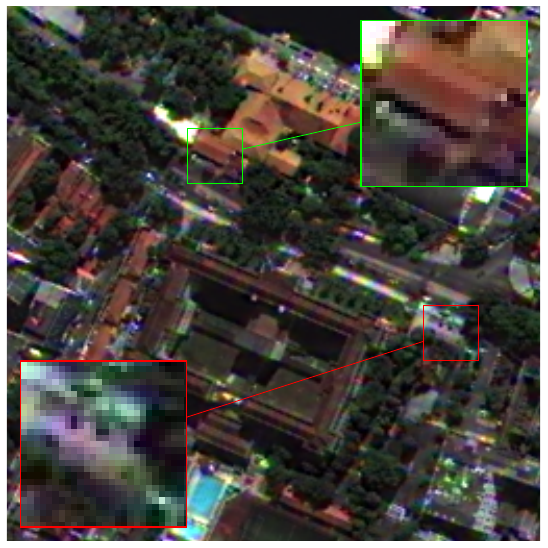}
		\caption{8 bands - Classic v2}
		\label{fig:janeiro_8_classic_v2}
	\end{subfigure}

	\caption[Band analysis for the \gls{mrca} image reconstruction]{Visual comparison of different \gls{mrca} image reconstruction algorithms, employing a different amount of embedded bands from the ``Janeiro'' dataset. In all cases, we visualize the \gls{rgb} bands of $256 \times 256$ \si{px} cropped area with zoomed details.}
	\label{fig:janeiro_bands}
\end{figure*}

    	The \gls{jodefu} algorithm achieves results which are at least comparable to the state-of-the-art.
    	The benefit of the \gls{tv}-style regularizer is immediately evident for the \gls{cassi} reconstruction (\figurename~\ref{fig:washington_4_cassi_jodefu}). For the pansharpening problem, the final product obtained with the \gls{jodefu} (\figurename~\ref{fig:washington_4_pansharpening_jodefu}) shows subtler improvements, as the final product achieves slightly more accurate color quality. For the demosaicing problem, however, there are still margins of improvement (\figurename~\ref{fig:washington_4_demosaic_jodefu}). The proposed algorithm is not fully capable of eliminating some texture effects; this is a known weakness of \gls{tv}-based regularizer, which are not well suited for the reconstruction of thread-like structures~\cite{Cham16}.
    	In the supplementary materials, we also present a visual comparison for the ``Fields'' dataset, where the terrain has a smoother structure compared to the more piecewise appearence of the ``Washington'' one. The \gls{mrca} reconstruction results are also compared to the image quality obtained with standard software compression algorithms aimed at matching the compression ratio achieved by the \gls{mrca}~\cite{Pico18b}.

    \subsection{Image reconstruction}
	\label{ssec:experiments_reconstruction}

		In this section, we shift the focus on image reconstruction methods, and compare the quality of the estimated products achieved with different methods, starting from an observation acquired with the full version of the proposed \gls{mrca} architecture. Our main target is to show the robustness of the proposed algorithm to recover the desired information when a sufficiently large amount of channels are embedded in the observation.

		The analysis is carried out by simulating the observation starting from the ``Janeiro'' and ``Stockholm'' dataset bundle. For the \gls{lri}, we select either 3, 4, or 8 channels, by either only choosing the \gls{rgb} in the first case, adding a \gls{nir} in the second one, or selecting all the \gls{vis} and \gls{nir} channels in the last case. The \gls{mrca} model employs the periodic masks shown in \figurename~\ref{fig:cfa_pattern}.

		Other than with \gls{jodefu}, this specific problem can be also approached with a custom image formation method, composed of the following three-step procedure:
		\begin{itemize}
			\item \textbf{\gls{hri} interpolation}: recovering the sparse channel associated to the \gls{hri} ($\mathbf{p}_{:,1}\masked=\mathbf{y}\hadam\mathbf{h}_{:,1}^{[p]}$) and estimate the missing \gls{hri} samples with a multivariate interpolation.
			\item \textbf{Demosaicing}: obtaining the \gls{lri} mosaic by decimating the observation $\mathbf{y}$. Furthermore, apply any classic demosaicing algorithm to estimate all the channels of the \gls{lri};
			\item \textbf{Sharpening}: performing a fusion on the reconstructed \gls{hri} and \gls{lri} from the previous two steps.
		\end{itemize}

		In our tests, we employ a \textit{\gls{tps-rbf}}~\cite{Buhm03} for the \gls{hri} interpolation. For a comprehensive comparison, we isolated two viable configurations for the demosaicing and pansharpening algorithms, whose specifics are given in \tablename~\ref{tab:jodefu_variants}:
		\begin{itemize}
			\item The \textbf{classic v1} setup is optimized for the \gls{rgb} setup and employs the \textit{\gls{ari}} demosaicing method~\cite{Monn17}, which is only applicable to Bayer masks.
			\item The \textbf{classic v2} method can be applied to all cases, and employs the \textit{\gls{gsa}} fusion method~\cite{Aiaz07}, as it provides more robust results for larger amount of bands.
		\end{itemize}
	
		\begin{figure*}
	\captionsetup[subfigure]{justification=centering}
	\centering
	\begin{subfigure}[t]{0.24\linewidth}
		\includegraphics[width=\linewidth]{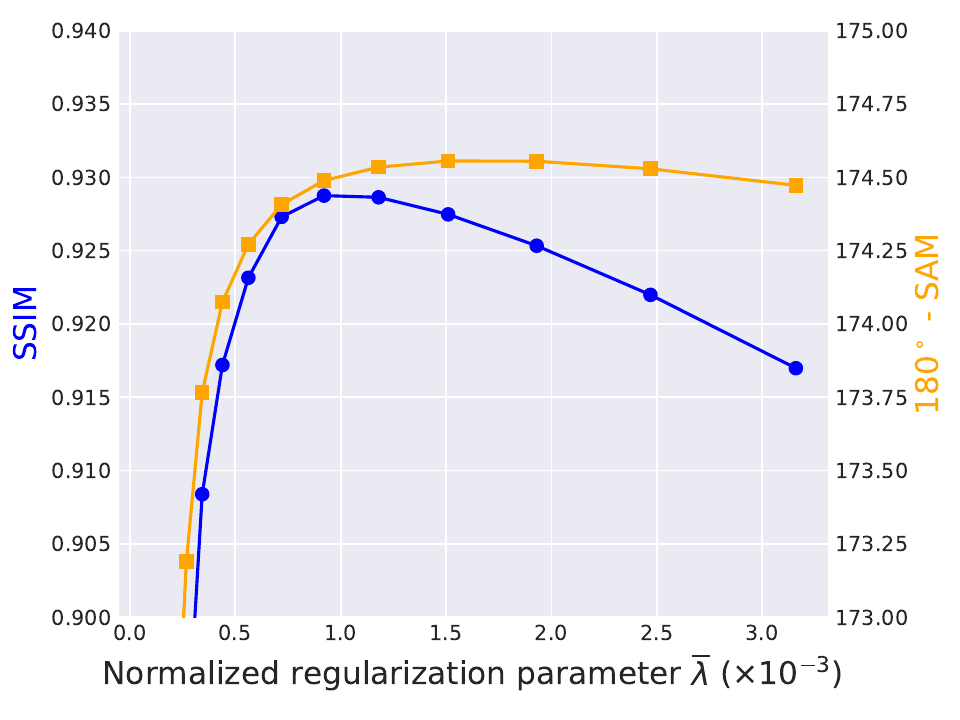}
		\caption{Regularization parameter $\lambda$}
		\label{fig:beijing_lambda}
	\end{subfigure}
	\hfil
	\begin{subfigure}[t]{0.24\linewidth}
		\includegraphics[width=\linewidth]{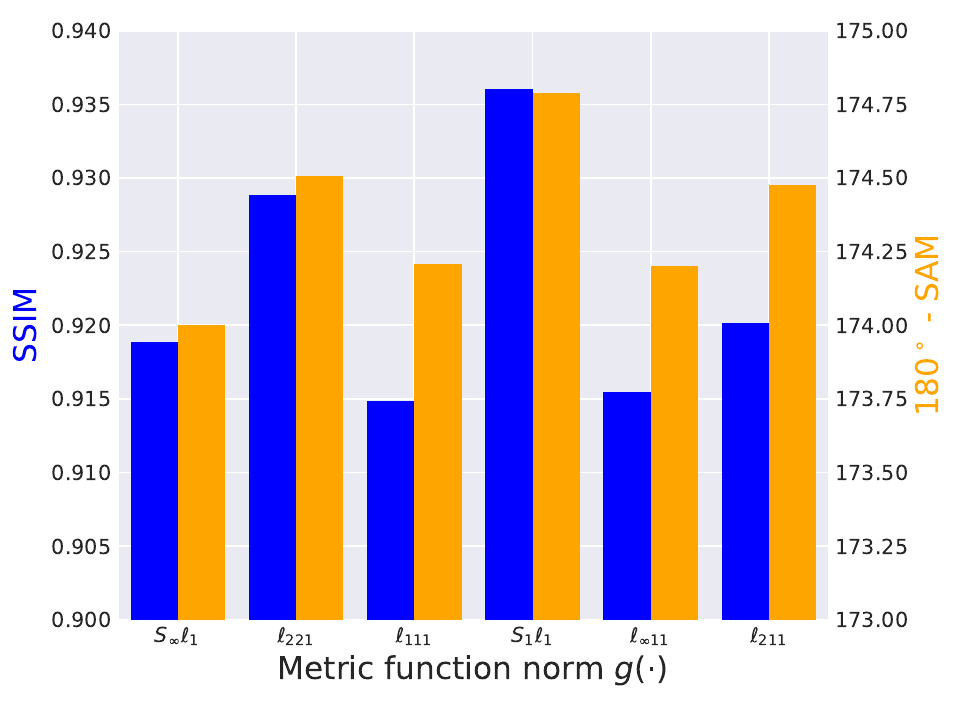}
		\caption{Metric operator norm $g(\cdot)$}
		\label{fig:beijing_norm}
	\end{subfigure}
	\hfil
	\begin{subfigure}[t]{0.24\linewidth}
		\includegraphics[width=\linewidth]{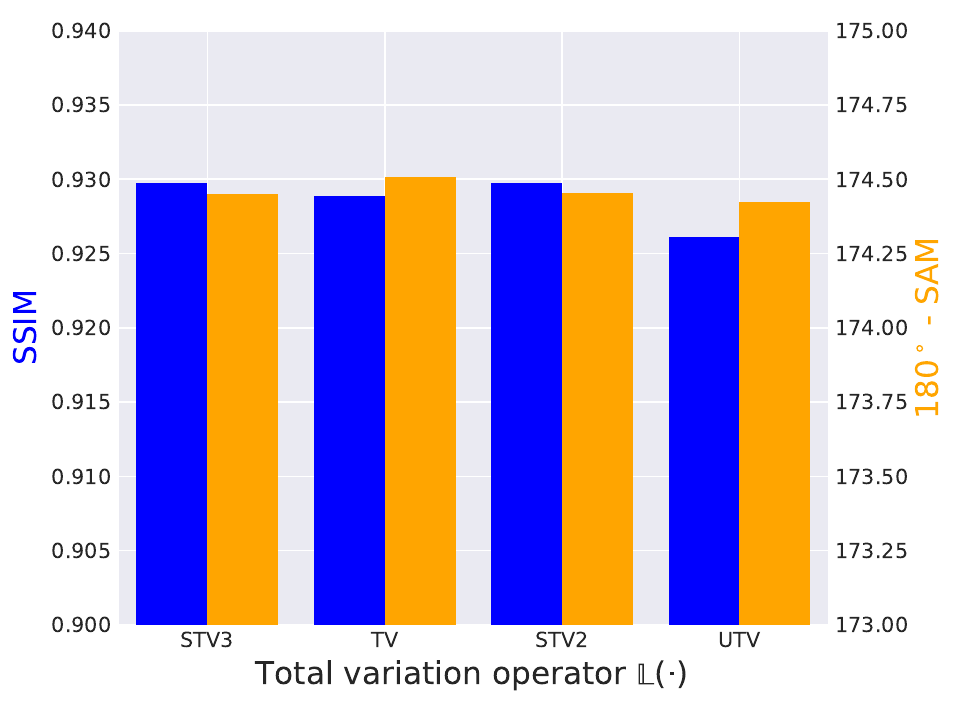}
		\caption{Total variation $\lino{L}(\cdot)$}
		\label{fig:beijing_tv}
	\end{subfigure}
	\hfil
	\begin{subfigure}[t]{0.24\linewidth}
		\includegraphics[width=\linewidth]{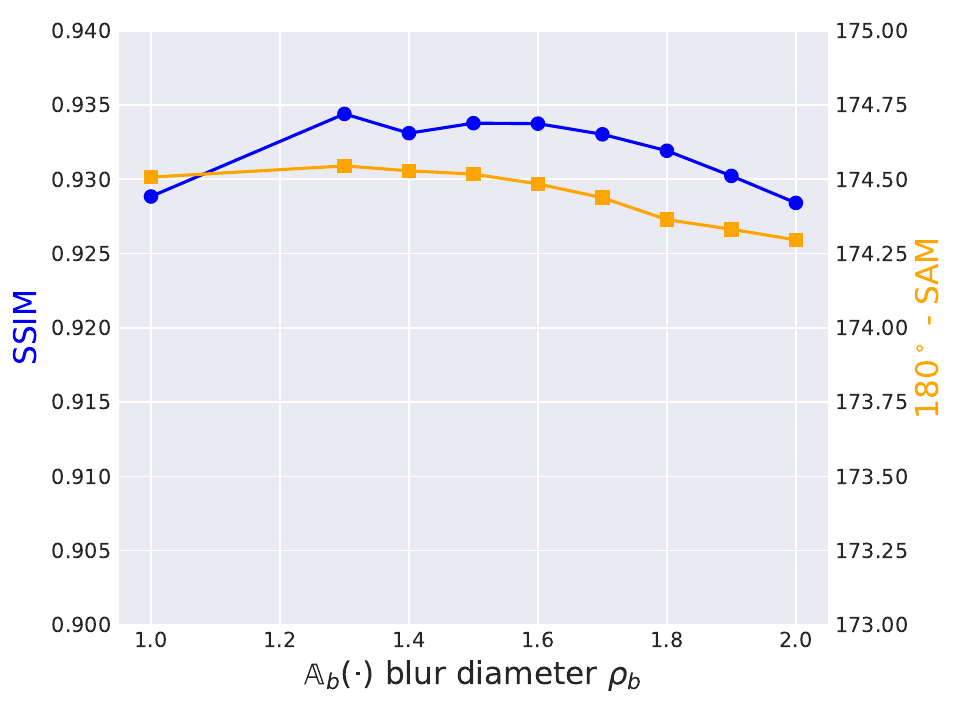}
		\caption{Blur diameter $\rho_b$ of $\lino{A}_b(\cdot)$}
		\label{fig:beijing_blur}
	\end{subfigure}
	
	\caption[Quality indices for different parameters' setups.]{Quality indices results obtained with different setups of the \gls{jodefu} image reconstruction algorithm applied to the 4-band ``Beijing'' dataset. For each figure, we vary the indicated parameter with respect to the baseline \gls{jodefu} v1 setup. Better reconstruction performances are associated to higher values of the ``\gls{ssim}'' (in blue) and of the ``$180^\circ -$ \gls{sam}'' (in orange).}
	\label{fig:beijing_parameters_graphs}
\end{figure*}

		The analysis of \tablename~\ref{tab:janeiro_bands} and its associated visual comparison of \figurename~\ref{fig:janeiro_bands} shows that the \gls{jodefu} yields sharper results than the classic v2, regardless of band setup. However, it does not provide a consistent reconstruction of larger scale homogeneous zones, such as the swimming pool in \figurename~\ref{fig:janeiro_4_jodefu}.
		This is most likely due to the choice the regularization parameter $\lambda$, which is not properly set to control the smoothing of zones with differents spatial scales, that would require to change the value of $\lambda$ locally.
		Some additional effects may come into play which are due to the nature of the datasets employed in this work, as remote sensing images are particularly sensitive to aliasing effects~\cite{Scho02}. Additionally, it is likely that a more accurate choice of the encoder may yield to more accurate reconstructions, as the regularity of the residual textures may be due to a sub-optimal choice of the mask coefficients.

    \subsection{Setting the parameters}
    \label{ssec:experiments_parameters}

\begin{figure*}
	\captionsetup[subfigure]{justification=centering}
	\centering
	\begin{subfigure}[t]{0.24\linewidth}
		\centering
		\includegraphics[width=\linewidth]{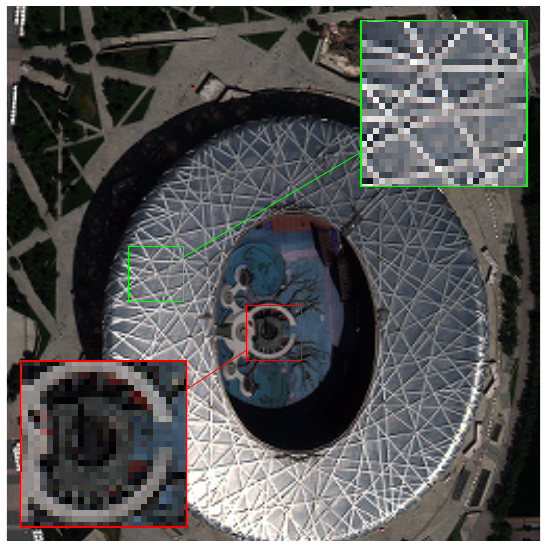}
		\caption{Reference (\glsentryshort{gt})}
		\label{fig:beijing_4_gt}
	\end{subfigure}
	\hfil
	\begin{subfigure}[t]{0.24\linewidth}
		\centering
		\includegraphics[width=\linewidth]{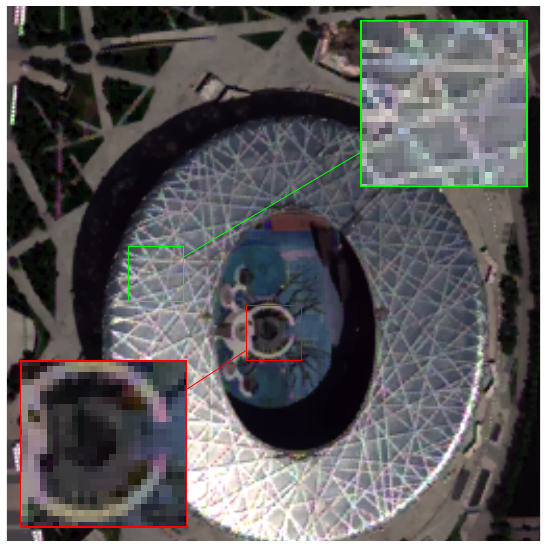}
		\caption{$g(\tens{W}) = \|\tens{W}\|_{S_1\ell_1}$}
		\label{fig:beijing_4_norm_S1l1}
	\end{subfigure}
	\hfil
	\begin{subfigure}[t]{0.24\linewidth}
		\centering
		\includegraphics[width=\linewidth]{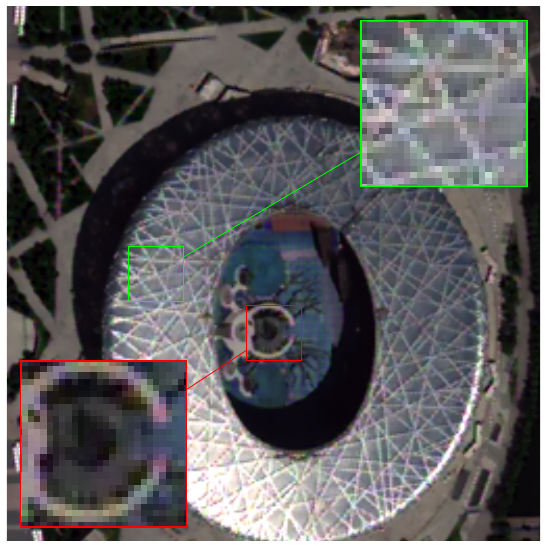}
		\caption{$\lino{L}(\cdot)$: \glsentryshort{stv} (upsc. ratio 2)}
		\label{fig:beijing_4_stv_2}
	\end{subfigure}
	\hfil
	\begin{subfigure}[t]{0.24\linewidth}
		\centering
		\includegraphics[width=\linewidth]{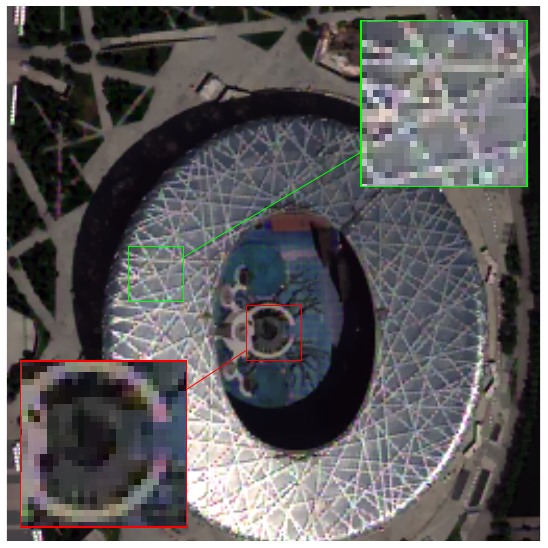}
		\caption{$\rho_b = 1.4$ \si{px}}
		\label{fig:beijing_4_blur_14}
	\end{subfigure}

	\smallskip
	
	\begin{subfigure}[t]{0.24\linewidth}
		\centering
		\includegraphics[width=\linewidth]{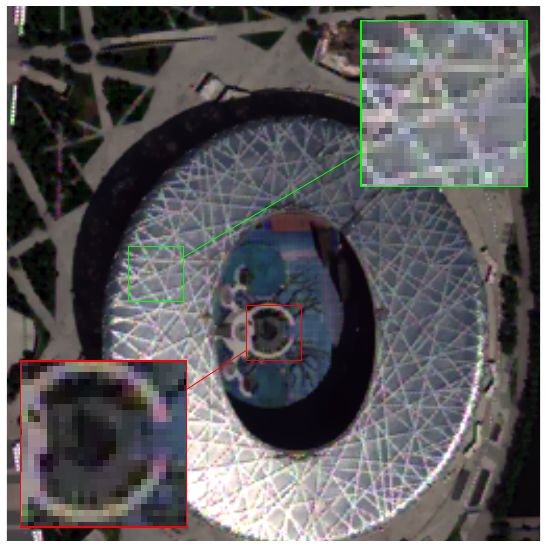}
		\caption{\glsentryshort{jodefu} v1 (baseline)}
		\label{fig:beijing_4_baseline}
	\end{subfigure}
	\hfil
	\begin{subfigure}[t]{0.24\linewidth}
		\centering
		\includegraphics[width=\linewidth]{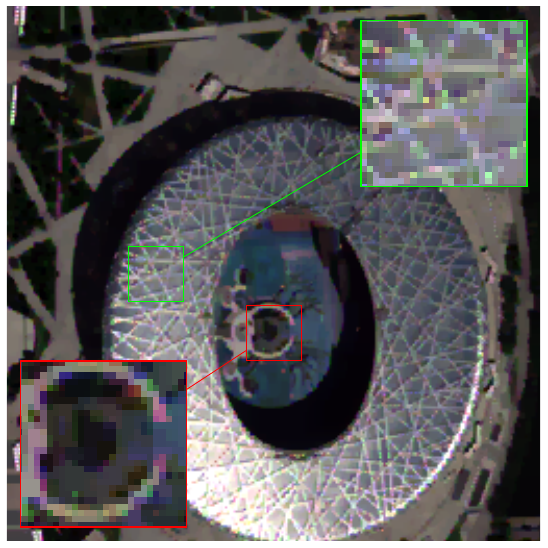}
		\caption{$g(\tens{W}) = \|\tens{W}\|_{111}$}
		\label{fig:beijing_4_norm_l111}
	\end{subfigure}
	\hfil
	\begin{subfigure}[t]{0.24\linewidth}
		\centering
		\includegraphics[width=\linewidth]{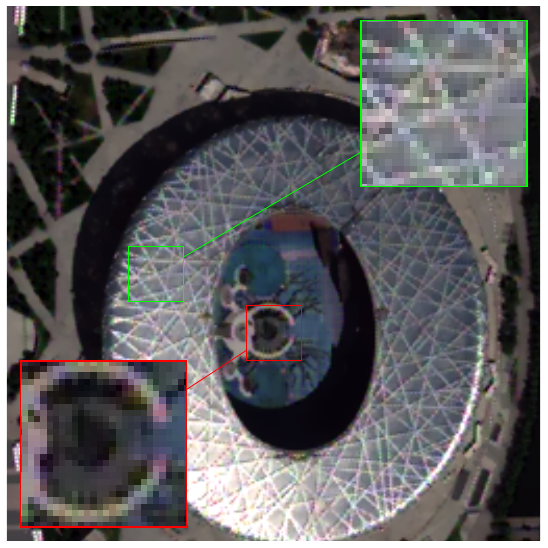}
		\caption{$\lino{L}(\cdot)$: \glsentryshort{utv}}
		\label{fig:beijing_4_tv_u}
	\end{subfigure}
	\hfil
	\begin{subfigure}[t]{0.24\linewidth}
		\centering
		\includegraphics[width=\linewidth]{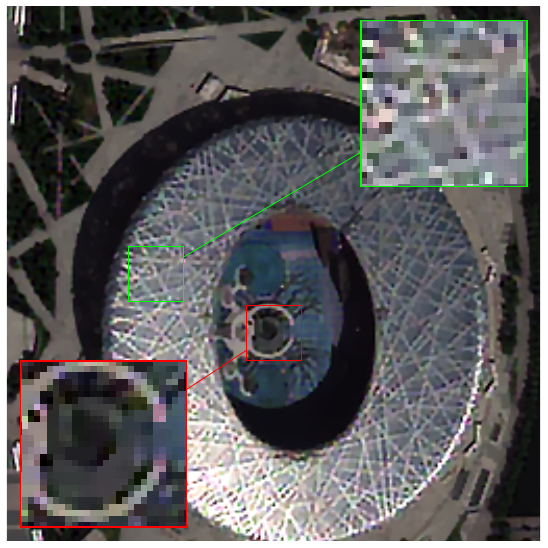}
		\caption{$\rho_b = 2.0$ \si{px}}
		\label{fig:beijing_4_blur_20}
	\end{subfigure}

	\smallskip
	\begin{subfigure}[t]{0.24\linewidth}
		\centering
		\includegraphics[width=\linewidth]{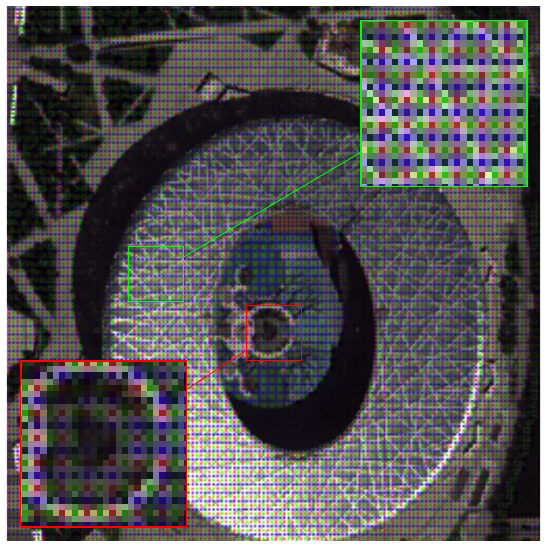}
		\caption{$\overline{\lambda}=1\times 10^{-4}$}
		\label{fig:beijing_4_lambda_min}
	\end{subfigure}
	\hfil
	\begin{subfigure}[t]{0.24\linewidth}
		\centering
		\includegraphics[width=\linewidth]{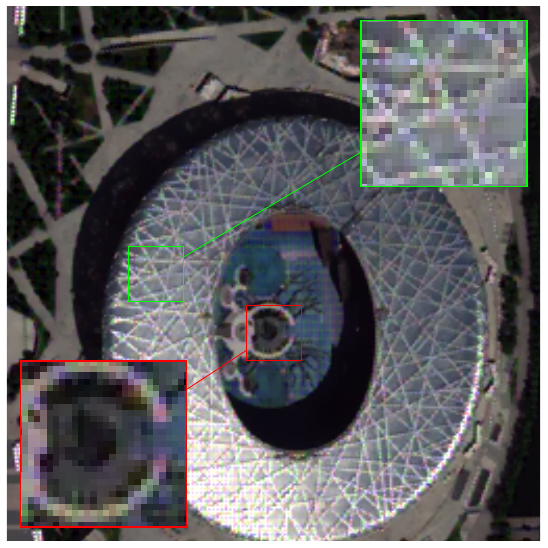}
		\caption{$\overline{\lambda}=6\times 10^{-4}$}
		\label{fig:beijing_4_lambda_mid_min}
	\end{subfigure}
	\begin{subfigure}[t]{0.24\linewidth}
		\centering
		\includegraphics[width=\linewidth]{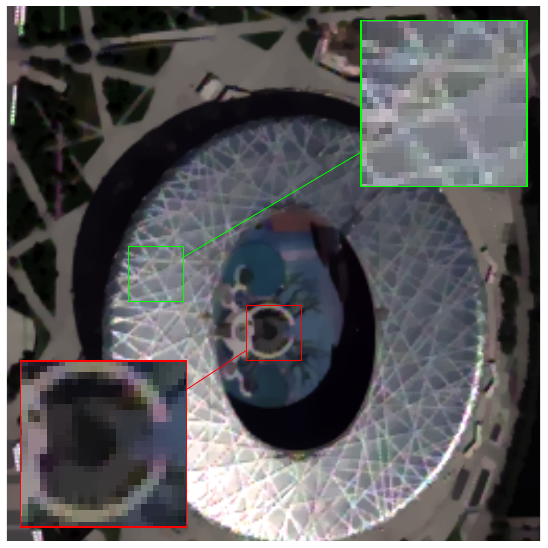}
		\caption{$\overline{\lambda}=7\times 10^{-3}$}
		\label{fig:beijing_4_lambda_mid_max}
	\end{subfigure}
	\hfil
	\begin{subfigure}[t]{0.24\linewidth}
		\centering
		\includegraphics[width=\linewidth]{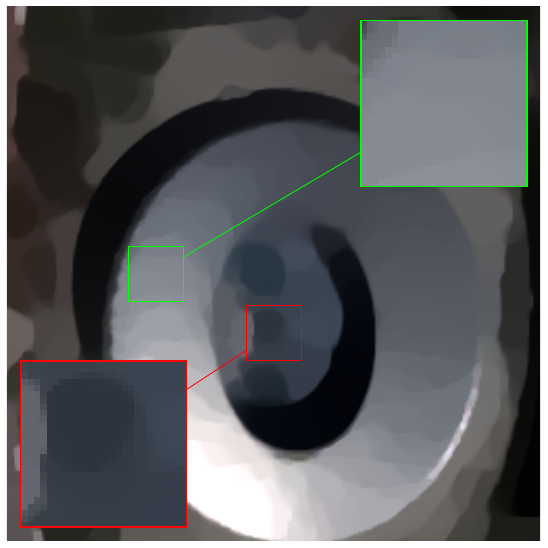}
		\caption{$\overline{\lambda}=1\times 10^{-1}$}
		\label{fig:beijing_4_lambda_max}
	\end{subfigure}
	
	\caption[Parameters visual comparison]{Visual comparison of the effects of changing parameters in the baseline setup of the \gls{jodefu} image reconstruction applied to the \gls{mrca} acquisition of the 4-band ``Beijing'' dataset ($256 \times 256$ \si{px} cropped area).}
	\label{fig:beijing_parameters}
\end{figure*}

        We test here various possible parameters for the optimization of the \gls{jodefu} algorithm applied to the \gls{mrca} acquisition obtained with the mask of \figurename~\ref{fig:cfa_pattern_ubt4}. For this test, we firstly define as \textit{baseline} the \gls{jodefu} v1 with $\overline{\lambda}=1\times10^{-3}$ (whose specifics are given in \tablename~\ref{tab:jodefu_variants}) and evaluate its estimated product. Then, for each parameter under test, we estimate the reconstruction results obtained by solely varying a single parameter from the baseline setup.
        There is no guarantee on the combined effect of varying multiple parameter, but we empirically experienced that optimizing each parameter separately still returns performances within a reasonable ballpark of the overall best optimization.
        The tests are applied to the $4$-band ``Beijing'' dataset, and a summary of the measured quality indices is given in \figurename~\ref{fig:beijing_parameters_graphs}, with an associated visual comparison in \figurename~\ref{fig:beijing_parameters}.

        A more in-depth discussion for each of the parameters under test is given in the following list:
        \begin{itemize}
            \item \textbf{Regularization Parameter} $\lambda$: as a rule of thumb, $\overline{\lambda}=10^{-3}$ is a good compromise in most scenarios and can be used as starting test to further refine the parameters if higher quality is required. In \figurename~\ref{fig:beijing_4_lambda_min} and \ref{fig:beijing_4_lambda_max}, some reconstructed products are shown for implausibly low and high values of $\lambda$, respectively; some inbetween choices are also shown for the sake of completion. If $\lambda$ is too low, we impose no structure of the final image, and most texture effects from the mosaicing are not flattened. If $\lambda$ is too high, the smoothing effect applies to relevant image features;

        	\item \textbf{Metric function norm} $g(\cdot)$: the quantitative verification shows that the $\ell_{221}$ norm is the best compromise between quality of the reconstructed product and computational speed. Among the remaining choices, better performances are only achieved with the $S_1\ell_1$, due to the noise whitening effect that this constraint imposes across different bands. The visual analysis shows some spectral spot-shaped spectral distortions in \figurename~\ref{fig:beijing_4_norm_l111};

        	\item \textbf{Linear operator} $\lino{L}(\cdot)$: for this configuration we tested the classic \gls{tv}, the \gls{utv}~\cite{Cham11} and the \gls{stv}~\cite{Aber17} with an upscaling factor of $2$ and $3$, but no noticeable differences were found;

            \item \textbf{Diameter} $\rho_b$ \textbf{of the blurring operator} $\lino{A}_b(\cdot)$: in our tests, the optimal value of the blur diameter was found to be in the range $1.3-1.5$ \si{px} for a scale ratio $\ratio=2$. This optimal value has to be chosen as a trade-off between a more accurate recovery of the \gls{hri} samples and avoiding out-of-focus effects in the final product.

        \end{itemize}

\section{Conclusion}
\label{sec:conclusions}

    In this paper, we proposed the \gls{mrca}, a novel multiresolution compressed acquisition system with enough flexibility to model both classic image formation methods, such as the \gls{cfa}/\gls{msfa}-based mosaicing, the \gls{hri}/\gls{lri} image bundles, and some nonconventional acquisition system, such as the \gls{cassi}, and various hybrid methods. The proposed design aims to intercept the future trends for unconventional optical devices, and allows for an immediately available procedure to model the physics of the image formation.

    We also proposed the \gls{jodefu}, a Bayesian solver for image reconstruction applicable to any available variant of \gls{mrca}-based acquisitions, and specialized on data with a strong \gls{lri} component. The proposed algorithm jointly addresses the problem of image fusion and reconstruction of compressed data, exploits the \gls{ctv} regularization to recover the desired product, and can be declined into two different variants with respect of the requirements in term of computation time. The proposed method does not necessarily match the state-of-the-art for every available image formation setup, but some alternatives were analyzed for particularly simple ones, which employ a cascade of classic techniques.

	A possible extension of this work may involve supervised learning~\cite{Laum22, Mong21} to fine tune the parameters of the \gls{jodefu} algorithm. Additionally, improved robustness to a higher number of channels can be obtained by capturing samples directly in a sparse domain, and the reconstructed products can achieve higher quality with suitable mask designs based on compressed sensing, expanding the results of our previous works~\cite{Pico18b, Pico21}.

\ifCLASSOPTIONcaptionsoff
  \newpage
\fi

\bibliographystyle{IEEEtran}
\bibliography{biblio.bib,biblio_local.bib}

\begin{IEEEbiography}[{\includegraphics[width=1in,height=1.25in,clip,keepaspectratio]{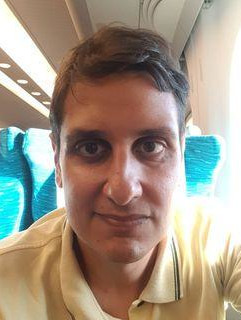}}]{Daniele Picone}(Member, IEEE)
	received the B.Sc. and M.Sc. degrees in electronic engineering from the University of Salerno, Salerno, Italy, in 2008 and 2016, respectively, and the Ph.D degree in Signal and Image processing form the University of Grenoble Alpes, Grenoble, France, in 2021.
	In 2016, he was a Research fellow for 4 months at University of Salerno. In 2019, he was a visiting researcher at the Tokyo Insitute of Technology, Japan for three months. Since December 2021, he was a Postdoctoral fellow for one year with the University of Grenoble Alpes, Grenoble, France, and then with the Grenoble Institute of Technology (Grenoble-INP), Grenoble, France, starting from January 2023.
	He is conducting his research at the Grenoble Images Speech Signals and Automatics Laboratory (GIPSA-Lab).
	His main research activities are in the fields of image processing and remote sensing, with applications mainly involving computational imaging, optimization, data fusion, and hyperspectral data processing.
\end{IEEEbiography}

\begin{IEEEbiography}[{\includegraphics[width=1in,height=1.25in,clip,keepaspectratio]{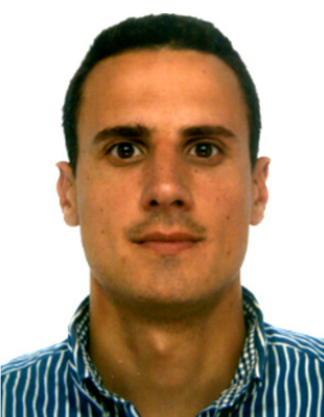}}]{Mauro Dalla~Mura}(Senior Member, IEEE)
	received the B.Sc. and M.Sc. degrees in Telecommunication Engineering from the University of Trento, Italy in 2005 and 2007, respectively.
	He obtained in 2011 a joint Ph.D. degree in Information and Communication Technologies (Telecommunications Area) from the University of Trento, Italy and in Electrical and Computer Engineering from the University of Iceland, Iceland.
	
	In 2011 he was a Research fellow at Fondazione Bruno Kessler, Trento, Italy, conducting research on computer vision.
	He is currently an Assistant Professor at Grenoble Institute of Technology (Grenoble INP), France since 2012.
	He is conducting his research at the Grenoble Images Speech Signals and Automatics Laboratory (GIPSA-Lab). He is a Junior member of the Institut Universitaire de France (2021-2026).
	Dr. Dalla Mura has been appointed "Specially Appointed Associate Professor" at the School of Computing, Tokyo Institute of Technology, Japan for 2019-2022.
	His main research activities are in the fields of remote sensing, computational imaging, image and signal processing.
	
	Dr. Dalla Mura was the recipient of the IEEE GRSS Second Prize in the Student Paper Competition of the 2011 IEEE IGARSS 2011 and co-recipient of the Best Paper Award of the International Journal of Image and Data Fusion for the year 2012-2013 and the Symposium Paper Award for IEEE IGARSS 2014.
	Dr. Dalla Mura was the IEEE GRSS Chapter's Committee Chair for 2020-2021. He was President of the IEEE GRSS French Chapter 2016-2020 (he previously served as Secretary 2013-2016). In 2017 the IEEE GRSS French Chapter was the recipient of the IEEE GRSS Chapter Award and the ``Chapter of the year 2017'' from the IEEE French Section.
	He is on the Editorial Board of the IEEE Journal of Selected Topics in Applied Earth Observations and Remote Sensing (J-STARS) since 2016.
\end{IEEEbiography}

\begin{IEEEbiography}[{\includegraphics[width=1in,height=1.25in,clip,keepaspectratio]{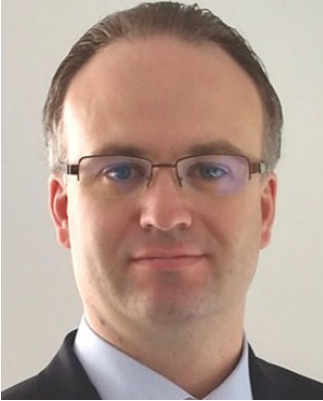}}]{Laurent Condat}
	(Senior Member, IEEE) received two Master's degrees in 2003, in computer science and applied mathematics, and a PhD in applied mathematics in 2006 from Grenoble Institute of Technology, Grenoble, France. After 2 years as a postdoc in the Helmholtz Zentrum Muenchen, Munich, Germany, he was hired as a permanent researcher by the French National Center for Scientific Research (CNRS). Since Nov. 2019, he is on leave from the CNRS and a Senior Research Scientist in King Abdullah University of Science and Technology (KAUST), Saudi Arabia.
	
	Dr. Condat's area of interest spans optimization, signal and image processing, inverse problems, and machine learning. He has co-authored more than 100 articles on these topics. He is a senior member of the IEEE and an associate editor of IEEE Transactions on Signal Processing (TSP). He received a best student paper award at the conference IEEE ICIP 2005, a best PhD award from Grenoble Institute of Technology in 2007, and several meritorious reviewer awards.
\end{IEEEbiography}

\end{document}